\documentclass[12pt, a4paper]{article}
\usepackage[dvipdfmx]{graphicx}
\usepackage{amssymb}
\usepackage{amsmath}
\usepackage{bm}
\usepackage{color}
\usepackage{cite}
\usepackage{slashed}
\usepackage{subfigure}
\usepackage{epstopdf}            
\usepackage{epsfig}
\usepackage{here}
            
\newcommand{\beq}{\begin{equation}}
\newcommand{\eeq}{\end{equation}}

\usepackage[colorlinks=true, linkcolor=black, citecolor=black,
urlcolor=black]{hyperref}

\begin{document}

\begin{titlepage}

\begin{center}

{\large
{\bf 
Status of the semileptonic $B$ decays and muon g-2  \\
in general 2HDMs with right-handed neutrinos}
}

\vskip 2cm

Syuhei Iguro$^{1}$
and
Yuji Omura$^{2}$

\vskip 0.5cm

{\it $^1$Department of Physics,
Nagoya University, Nagoya 464-8602, Japan}\\[3pt]

{\it $^2$
Kobayashi-Maskawa Institute for the Origin of Particles and the
Universe, \\ Nagoya University, Nagoya 464-8602, Japan}\\[3pt]

\
\vskip 1.5cm

\begin{abstract}
In this paper, we study the extended Standard Model (SM) with an extra Higgs doublet and right-handed neutrinos.
If the symmetry to distinguish the two Higgs doublets is not assigned,
flavor changing neutral currents (FCNCs) involving the scalars are predicted even at the tree level. We investigate the constraints on the FCNCs at the one-loop level, and especially study the semileptonic $B$ meson decays, e.g.
$B \to D^{(*)} \tau \nu$ and $B \to K^{(*)} ll$ processes, where the SM predictions are more than $2 \sigma$ away
from the experimental results.
We also consider the flavor-violating couplings involving right-handed neutrinos and
discuss if the parameters to explain the excesses of the semileptonic $B$ decays
can resolve the discrepancy in the anomalous muon magnetic moment.
Based on the analysis, we propose the smoking-gun signals of our model at the LHC.
\end{abstract}

\end{center}
\end{titlepage}

\section{Introduction}
The Standard Model (SM) succeeds in describing almost all of the experimental results.
There is one Higgs doublet to break the electroweak (EW) symmetry,
and the non-vanishing vacuum expectation value (VEV) of the Higgs field 
generates the masses of the gauge bosons and the fermions.
We do not still understand the reasons why the EW scale is around a few hundred GeV
and why the couplings between the Higgs field and the fermions are so hierarchical.
The Higgs particle is, however, discovered at the LHC experiment, and
the signal is consistent with the SM prediction~\cite{Aad:2012tfa,Chatrchyan:2012xdj}. 
Thus, we are certain that the SM describes our nature up to the EW scale.

On the other hand, it would be true that the structure of the SM is so mysterious.
In addition to the mystery of the origin of the Higgs potential and couplings,
the structure of the gauge symmetry is also very non-trivial. 
The anomaly-free conditions are miraculously satisfied:
it is not easy to add extra chiral fermions to the SM. 
In the bottom-up approach to the new physics, 
one possible extension is to add extra scalars, e.g. extra Higgs doublets, to avoid the inconsistency with
the anomaly-free conditions. Such a simple extension opens up rich phenomenology, so that
a simple extended SM with an extra Higgs doublet has been actually discussed since about 40 years ago \cite{Lee:1973iz,Haber:1978jt,Donoghue:1978cj,Hall:1981bc,Hou:1991un,Chang:1993kw,Atwood:1995ud,Atwood:1996vj}.

The extended SM, besides, has other interesting aspects, from the viewpoint of the top-down approach.
If we consider the new physics that can solve the mysteries of the SM,
we often find extra Higgs doublets. For instance, the supersymmetric extension predicts at least one more
Higgs doublet. If we consider the extended gauge symmetry, such as $SU(2)_R$, we find
extra Higgs doublets that couple to the SM fermions in the effective lagrangian. If we assume that
there are flavor symmetries at high energy, there would be many Higgs doublets that couple
to the SM fermions flavor-dependently.
Thus, it would be very interesting and important to study and summarize
the predictions and the experimental constraints of the extended SM with extra Higgs doublets.

Based on this background, we investigate not only the experimental constraints but also
the predictions for the observables relevant to the future experiments, in the extended SM with one Higgs doublet (2HDM).
We adopt the bottom-up approach.
In our model, we do not assign any symmetry to distinguish the two Higgs doublets, so that
there are tree-level Flavor Changing Neutral Currents (FCNCs) involving scalars \cite{Glashow:1976nt}.
This kind of general 2HDM has been discussed, and often called the Type-III 2HDM \cite{Hou:1991un,Chang:1993kw,Atwood:1996vj,Liu:1987ng,Cheng:1987rs,Savage:1991qh,Antaramian:1992ya,Hall:1993ca,Luke:1993cy,Atwood:1995ud,Aoki:2009ha}. Hereafter, we abbreviate such a generic 2HDM with tree-level FCNCs as the Type-III 2HDM.
We note that this kind of setup is predicted as the effective model of 
the extended SM with the extended gauge symmetry; e.g., the left-right symmetric model \cite{Haba:2017jgf} and the SO(10) grand unified theory \cite{progress}. In our model, we also introduce right-handed neutrinos and allow the coupling between the right-handed neutrinos and both Higgs doublets.
We simply assume that the light neutrinos are Dirac fermions, and 
the tiny masses are given by the small Yukawa couplings.
Although the fine-tuning may be required, the Yukawa couplings between the neutrino and the extra scalars
could be sizable in principle.\footnote{We note that the right-handed neutrino can have the Majorana mass term. Our discussion, however, does not change, as far as the Majorana mass is small and it is irrelevant to the active neutrino.}

Recently, the Type-III 2HDM is attracting a lot of attention, since it is
one of the good candidates to explain the excesses reported by the BaBar,  Belle, and LHCb collaborations. 
In the experiments, the semileptonic $B$ decays, $B \to D^{(*)} \tau \nu$, have been measured 
and the results largely deviate from the SM predictions \cite{Lees2012xj,Lees2013uzd,Huschle2015rga,Sato2016svk,Hirose2016wfn,Aaij2015yra,FPCP2017LHCb,Jaiswal:2017rve}.
The $B$ decays in the Type-III 2HDM have been studied in Refs. \cite{Ko:2017lzd,Ko:2012sv,Iguro:2017ysu,Crivellin2012ye,Celis:2012dk,Tanaka2012nw,Crivellin:2013wna,Crivellin:2015hha,Cline:2015lqp,
Hu:2016gpe,Arnan:2017lxi,Arhrib:2017yby,Bian:2017rpg,Bian:2017xzg}.
Although we recently find that the explanation of $B \to D^{*} \tau \nu$ contradicts 
the leptonic $B_c$ decay \cite{Alonso:2016oyd,Akeroyd:2017mhr}, the Type-III 2HDM is still one of the plausible and attractive candidates
to achieve the explanation of the excess in $B \to D \tau \nu$ \cite{Iguro:2017ysu}.
In addition, another semileptonic $B$ decay, i.e. $B \to K^{(*)} \mu \mu$, is also discussed recently in the 2HDM \cite{Hu:2016gpe,Arnan:2017lxi,Arhrib:2017yby}. In the process, the LHCb collaboration has reported the deviations
from the SM predictions in the measurements concerned with the angular observables \cite{Aaij:2013qta,Aaij:2015oid} and the lepton universality \cite{LHCbnew,Aaij:2014ora}.
Moreover, it is known that the Type-III 2HDM can accomplish the explanation of the anomalous muon magnetic moment ($(g-2)_\mu$) deviated from the SM prediction \cite{Omura:2015nja,Omura:2015xcg}.

In fact, the each explanation is elaborately achieved by tuning some proper parameters, since
the experimental constraints are very strong in all cases.
There are many parameters in the Type-III 2HDM, so that it may be possible to find
a parameter set to explain the all excesses.
In this paper, we discuss the compatibility between each of the explanations.
Compared to the previous works \cite{Iguro:2017ysu,
Hu:2016gpe,Arnan:2017lxi,Arhrib:2017yby}, 
we take into consideration the constraint from the lepton universality of $B \to D^{(*)} l \nu$ $(l=e,\, \mu)$.
The compatibility of those excesses in the $B$ decays with the $(g-2)_\mu$ discrepancy has not been also studied before.
We also consider the contributions of the flavor violating couplings involving the right-handed neutrinos.

This paper is organized as follows. In Sec. \ref{sec;setup}, we introduce our model and the simplified setup
to evade the strong experimental constraints. In Sec. \ref{sec;flavor}, we summarize the experimental constraints
on our model and discuss (semi)leptonic $B$ decays in the Type-III 2HDM in Sec. \ref{sec;flavor2}.
We also propose our signals at the LHC in Sec. \ref{sec;collider}.
Sec. \ref{sec;summary} is devoted to the summary.

\section{Type-III 2HDM}
\label{sec;setup}
We introduce the Type-III 2HDM with right-handed neutrinos.
There are two Higgs doublets in our model. When the Higgs fields are written in the basis where only one Higgs doublet
obtains the nonzero VEV, the fields can be decomposed as \cite{Davidson:2005cw}
\begin{eqnarray}
  H_1 =\left(
  \begin{array}{c}
    G^+\\
    \frac{v+\phi_1+iG}{\sqrt{2}}
  \end{array}
  \right),~~~
  H_2=\left(
  \begin{array}{c}
    H^+\\
    \frac{\phi_2+iA}{\sqrt{2}}
  \end{array}
  \right),
\label{HiggsBasis}
\end{eqnarray}
where $G^+$ and $G$ are Nambu-Goldstone bosons, and $H^+$ and $A$ are a charged Higgs boson and a CP-odd
Higgs boson, respectively. $v$ is the VEV: $v \simeq 246$ GeV.
In this base, we write down the Yukawa couplings with the SM fermions.
In the mass basis of the fermions, the Yukawa interactions are expressed by \cite{Davidson:2005cw}
\begin{eqnarray}
  {\cal L}&=&-\bar{Q}_L^i H_1 y^i_d d_R^i -\bar{Q}_L^i H_2 \rho^{ij}_d d_R^j
  -\bar{Q}_L^i (V^\dagger)^{ij}\widetilde H_1 y^j_u u_R^j -\bar{Q}_L^i
  (V^\dagger)^{ij}\widetilde H_2
  \rho^{jk}_u u_R^k \nonumber\\
  &&-\bar{L}_L^i H_1 y^i_e e_R^i -\bar{L}_L^i H_2 \rho^{ij}_e e_R^j-\bar{L}_L^i (V_{\nu})^{ij} \widetilde H_1 y^j_\nu \nu_R^i -\bar{L}_L^i (V_{\nu})^{ij} \widetilde H_2 \rho^{jk}_\nu \nu_R^k,
\end{eqnarray}
where $i$, $j$ and $k$ represent flavor indices, and $Q=(V^\dagger u_L,d_L)^T$,
$L_L=(V_{\nu} \nu_L, e_L)^T$ are defined. 
$\widetilde H_{1,2}$ denote $\widetilde H_{1,2}=i \tau_2 H^*_{1,2}$, where $\tau_2$ is the Pauli matrix.
$V$ is the Cabbibo-Kobayashi-Maskawa (CKM) matrix
and $V_{\nu}$ is the Maki-Nakagawa-Sakata (MNS) matrix.
Fermions $(f_L,~f_R)$ $(f=u,~d,~e,~\nu)$ are mass eigenstates, and 
$ y_i^f=\sqrt{2}m_{f_i}/v$, where $m_{f_i}$ denote the fermion masses, are defined.
$\rho_f^{ij}$ are the Yukawa couplings that are independent of the SM fermion mass matrices.

There are three types of the scalars: the charged Higgs ($H^\pm$), the CP-odd scalar ($A$) and
the two CP-even scalars ($\phi_{1,2}$). The CP-even scalars are not mass eigenstates, although
the mixing should be tiny not to disturb the SM prediction. The mixing is defined as
\beq
\begin{pmatrix}\phi_1 \\ \phi_2 \end{pmatrix} = \begin{pmatrix} \cos \theta_{\beta \alpha} & \sin \theta_{\beta \alpha}   \\ -\sin \theta_{\beta \alpha} & \cos \theta_{\beta \alpha} \end{pmatrix}
\begin{pmatrix}h \\ H \end{pmatrix}. 
\eeq

The masses of the heavy scalars can be evaluated as
\begin{eqnarray}
  m_H^2& \simeq& m_A^2+\lambda_5 v^2, \\
  m_{H^\pm}^2 &\simeq& m_A^2-\frac{\lambda_4-\lambda_5}{2} v^2.
  \label{Higgs_spectrum}
\end{eqnarray}
$m_H$, $m_A$ and $m_{H^+}$ denote the masses of the heavy CP-even, CP-odd and charged Higgs scalars.
$\lambda_4$ and $\lambda_5$ are the dimensionless couplings in the Higgs potential:
$V(H_i)=\lambda_4 (H_1^\dagger H_2)(H_2^\dagger H_1) + \frac{\lambda_5}{2}(H_1^\dagger H_2)^2+ \dots $
The mass differences are relevant to the electro-weak precision observables (EWPOs) and the explanation of the $(g-2)_\mu$ anomaly \cite{Omura:2015nja,Omura:2015xcg}.

\subsection{Setup of the texture}
$\rho_f$ are $3\times 3$ matrices and the each element is the free parameter that is constrained by the flavor physics and the collider experiments.
The comprehensive study about the phenomenology in the Type-III 2HDM has been done in Ref. \cite{Crivellin:2013wna}.
There are many choices for the matrix alignment, but actually 
only a few elements are allowed to be sizable according to the stringent experimental bounds \cite{Crivellin:2013wna}.

First, let us discuss the physics concerned with $\rho_u$ and $\rho_d$.  
The all off-diagonal elements of $\rho_d$ are strongly constrained by the
$\Delta F=2$ processes. $\rho_u^{uc}$ and $\rho_u^{cu}$ have to be small to avoid
the stringent constraint that comes from the $D-\overline{D}$ mixing.
Besides, we find that the size of the Yukawa coupling involving the light quarks
are limited by the direct search at the collider experiments.
 Even $\rho_u^{ut}$ and $\rho_u^{tu}$ may be constrained by the bounds from the collider experiment, e.g.,
the upper limit from the same-sign top signal.\footnote{Note that there is a way to avoid the strong constraint, considering the degenerate masses of the scalars \cite{Ko:2011vd,Ko:2011di}. } 
Moreover, $\rho_u^{ut}$ and $\rho_u^{tu}$ are strongly constrained by the $K$-$\overline{K}$ mixing at the one-loop level.
Thus, it is difficult to
expect that the couplings between the light quarks ($u, \, d, \,s$) and the other quarks are larger than ${\cal O} (0.01)$.
The diagonal elements, on the other hand, could be ${\cal O} (0.1)$, unless
the off-diagonal elements are not sizable \cite{Omura:2015nwa}.

Based on the examination, we consider the case that $|\rho_u^{ct}|$ and/or $|\rho_u^{tc}|$ are sizable.
One of our motivations of this study is to investigate the compatibility among the 
explanations of the excesses in the Type-III 2HDM.
It is pointed out that
the sizable $\rho_u^{tc}$  can improve the discrepancies in the $b \to s ll$ and $b \to c l \nu$ processes \cite{Iguro:2017ysu}. Eventually, we consider the following simple textures of $\rho_f$ from the phenomenological point of view: 
\begin{equation}
\label{eq;rhou}
\rho_u \simeq \begin{pmatrix} 0 & 0 & 0 \\ 0 & 0 & \rho_u^{ct} \\ 0 &   \rho_u^{tc} & \rho_u^{tt}  \end{pmatrix},~|\rho_d^{ij}| \ll {\cal O} (0.1).
\end{equation}
The other elements of $\rho_u$ are assumed to be at most ${\cal O} (0.01)$,
so that the physics involving $\rho_u^{ct}$, $\rho_u^{tc}$, and $\rho_u^{tt}$ is mainly discussed in this paper.
Note that we ignore all elements of $\rho_d$ and assume that all sizable Yukawa couplings
are real, through our paper.

Next, we discuss the Yukawa couplings with leptons.
We can also find the strong upper bounds on the Yukawa couplings in the lepton sector.
The lepton flavor violating (LFV) processes are predicted by the neutral scalar exchanging,
if the off-diagonal elements of $\rho_e$ are sizable.
In the case that the extra Yukawa couplings involving electron are large, the LEP experiment can 
easily exclude our model. Interestingly, the authors of Refs. \cite{Omura:2015nja,Omura:2015xcg}
have pointed out that the large $\rho_e^{\mu \tau}$ and $\rho_e^{\tau \mu}$
can achieve the explanation of the $(g-2)_{\mu}$,
that is largely deviated from the SM prediction.
The explanation, however, requires the other Yukawa couplings to be small \cite{Omura:2015nja,Omura:2015xcg}.
Then, we especially consider the following texture of $\rho_e$:
\begin{equation}
\label{eq;rhoe}
\rho_e \simeq \begin{pmatrix} 0 & 0 & 0 \\ 0 & 0 & \rho_e^{\mu \tau} \\ 0 &   \rho_e^{ \tau \mu} & 0  \end{pmatrix}.
\end{equation}
Note that the diagonal elements, $\rho_e^{ \tau \tau}$ and $\rho_e^{ \mu \mu}$, are also strongly constrained,
as far as $\rho_e^{\mu \tau}$ and $\rho_e^{\tau \mu}$ are sizable \cite{Omura:2015xcg}.

In our study, we also consider the contribution of $\rho_\nu$ to flavor physics.
This investigation has not been done well in the type-III 2HDM.
This is because the tiny Dirac neutrino masses predict small Yukawa couplings so that
$\rho_\nu$ is also naively expected to be small. 
$\rho_\nu$, however, does not contribute to the active neutrino masses, directly.
If both $\rho_\nu$ and $\rho_e$ are sizable, $\rho_\nu$ would contribute to the
neutrino masses radiatively. Otherwise, $\rho_\nu$ could be large compared to
$y^i_{\nu}$, in the bottom-up approach. 
The unique texture as in Eq. (\ref{eq;rhoe}) may also allow $\rho_\nu$ to be sizable. 
Based on this consideration, we study the upper bound on $\rho_\nu$ and
discuss the impact on the physical observables in flavor physics.


\section{The summary of the experimental constraints}
\label{sec;flavor}
In this section, we discuss the physics triggered by
the Yukawa couplings in Eq. (\ref{eq;rhou}) and Eq. (\ref{eq;rhoe}). 
The contribution of $\rho_\nu$ is also studied.
Note that we are interested in the light scalar scenario.
In order to avoid the exotic decay, e.g. $t \to H c$, and enlarge the new physics contributions maximumly, 
the extra scalar masses are set to 200 GeV or 250 GeV below.


\subsection{The experimental constraints on $\rho_u$ }
\label{sec;rhou}
To begin with, we summarize the experimental constraints on $\rho_u$.
In our study, the texture of $\rho_u$ is approximately given by Eq. (\ref{eq;rhou}).
Then, we can evade the strong bound from the $\Delta F=2$ processes at the tree level.
The measurements of the meson mixings are, however, very sensitive to new physics contributions,
so that we need to study the bounds carefully, taking into account the loop corrections.
\begin{figure}[t]
  \begin{center}
    \includegraphics[width=0.4\textwidth]{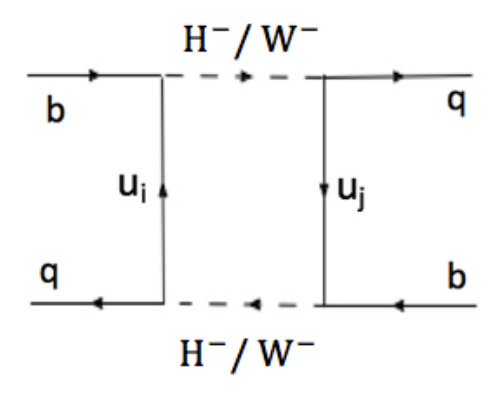}
    \caption{The diagrams that contribute to the $B_{(s)}-\overline{B_{(s)}}$ mixing.}
    \label{diagram-BBbar}
  \end{center}
\end{figure}

In our setup, the one-loop corrections involving the charged Higgs and the $W$-boson,
that are described in Fig. \ref{diagram-BBbar},
contribute to the $B$-$\overline{B}$ mixing and the $B_s$-$\overline{B_s}$ mixing.
The operators induced by the one-loop corrections are
\begin{align}
  {\cal H}^{\Delta F=2}_{\rm eff} &=-C_{LL}^q (\bar{q} \gamma^\mu P_L b)(\bar{q} \gamma_\mu P_L b),
\end{align}
where $q=s$, $d$.
The new physics contribution to the coefficient, $C_{LL}$, is evaluated at the one-loop level as
\begin{align}
 C^q_{LL}&=\frac{1}{128\pi^2 m_{H^+}^2}
    \sum_{k,l} (V^\dagger \rho_u)^{q k} (\rho^\dagger_u V)^{lb}\left[
      (\rho_u^\dagger V)^{kb}(V^\dagger \rho_u)^{q l} G_1(x_k,x_l) \right.\nonumber \\
      &\left.-\frac{4g^2m_{u_k} m_{u_l} }{m_{H^+}^2} V_{k b}V^*_{l q}G_2(x_k,x_l,x_W)
      +\frac{g^2 m_{u_k}m_{u_l}}{m_W^2} V_{k b} V^*_{l q}G_3(x_k,x_l,x_W)
      \right],
\end{align}
where $x_k=m_{u_k}^2/m_{H^+}^2$ and $x_W=m_W^2/m_{H^+}^2$.
The functions $G_i~(i=1,2,3)$ are defined as
\begin{align}
  G_1(x,y) &=\frac{1}{x-y}\left[
    \frac{x^2\log x}{(1-x)^2}+\frac{1}{1-x}-
    \frac{y^2\log y}{(1-y)^2}-\frac{1}{1-y}
    \right],
\label{G1_func}
  \\
  G_2(x,y,z) &=-\frac{1}{(x-y)(1-z)}
  \left[
    \frac{x \log x}{1-x}-\frac{y\log y}{1-y}-\frac{x\log\frac{x}{z}}{z-x}
    +\frac{y\log \frac{y}{z}}{z-y}
    \right],
  \\
  G_3(x,y,z) &=-\frac{1}{x-y}
  \left[
    \frac{1}{1-z}\left(\frac{x\log x}{1-x}-\frac{y\log y}{1-y}\right)
    -\frac{z}{1-z}\left(
    \frac{x\log \frac{x}{z}}{z-x}-\frac{y\log \frac{y}{z}}{z-y}\right)\right].
\end{align}
Using the coefficient, the mass difference, $\Delta m_{B_{d,s}}$, can be evaluated as 
\begin{equation}
  \Delta m_{B_{d_i}} =-2 {\rm Re} (C^q_{LL})\frac{m_{B_{d_i}} F_{B_{d_i}}^2 B_{B_{d_i}}}{3},
\end{equation}
where $m_{B_{d_i}}$, $F_{B_{d_i}}$ and $B_{B_{d_I}}$ are a mass, a decay constant and the bag parameter
of $B_{d_i}$ meson, respectively. We note that $C^q_{LL}$ includes the SM correction.

The deviations of the neutral $B_{(s)}$ meson mixing will be evaluated including the SM corrections, but it is certain that
there are non-negligible uncertainties in the theoretical predictions. 
In our analysis, we calculate our predictions, using the input parameters in Appendix \ref{input}.
In order to draw the constraints on the Yukawa couplings, we require that the deviations induced by the charged Higgs contributions are within the $2 \sigma$ errors of the SM predictions and the experimental results.
We simply adopt the SM predictions ($\Delta M_{B_{(s)}}^{SM}$) given by Ref. \cite{Bobeth:2016llm}:
$0.45~[{\rm ps}^{-1}] \le \Delta M_{B}^{\rm SM} \le 0.78~[{\rm ps}^{-1}]$ and $16.2~[{\rm ps}^{-1}]\le \Delta M_{B_s}^{\rm SM}\le21.9$ (95\% CL).
Then, we define $\delta (\Delta M_{B_{(s)}})= \Delta M_{B_{(s)}}^{exp}-\Delta M_{B_{(s)}}^{SM}$,
where $\Delta M_{B_{(s)}}^{exp}$ are the experimental values: $\Delta M_{B}^{exp}=0.5064\pm 0.0019~[{\rm ps}^{-1}]$ and $\Delta M_{B_{s}}^{exp}=17.757\pm 0.021~[{\rm ps}^{-1}]$ \cite{Amhis:2016xyh}.  
Taking into account the $2 \sigma$ uncertainties, $\delta (\Delta M_{B_{(s)}})$ are within the following ranges:
\beq
\label{bound;BBbar}
-0.27\le\delta(\Delta M_B) [\text{ps}^{-1}]\le0.06, ~ -4.1 \le\delta(\Delta M_{B_s}) [\text{ps}^{-1}]\le1.6.
\eeq
If the magnitudes of the Yukawa couplings are below the upper bounds in Table \ref{table1} when $m_{H^\pm}=200$ GeV and $250$ GeV, 
the charged Higgs contributions are within these ranges in Eq. (\ref{bound;BBbar}).
The results in Table \ref{table1} are consistent with the ones in Ref. \cite{Altunkaynak:2015twa}.
\begin{table}[H]
  \begin{center}
    \begin{tabular}{|c|c|c|c|}
      \hline
\multicolumn{4}{|c|}{  $B-\overline{B}$ Mixing}\\
\hline \hline
$m_{H^\pm}$&~~~$|\rho^{ct}_u|$~~~&~~~$|\rho^{tc}_u|$~~~&~~~$|\rho^{tt}_u|$~~~\\
\hline
200 [GeV]&0.307&1.00&0.741\\
250 [GeV]&0.340&1.12&0.814\\
\hline \hline
 \multicolumn{4}{|c|}{  $B_s-\overline{B_s}$ Mixing} \\
\hline \hline
$m_{H^\pm}$&$|\rho^{ct}_u|$&$|\rho^{tc}_u|$&$|\rho^{tt}_u|$\\
\hline
200 [GeV]&0.276&0.748&0.428\\
250 [GeV]&0.307&0.836&0.473\\
\hline
    \end{tabular}
    \caption{The upper bounds on the up-type Yukawa couplings from the $\Delta F=2$ processes, fixing $m_{H^\pm}$ at
    $m_{H^\pm}=200$ GeV and $250$ GeV.  }
    \label{table1}
  \end{center}
\end{table}

\begin{figure}[t]
  \begin{center}
    \includegraphics[width=0.4\textwidth]{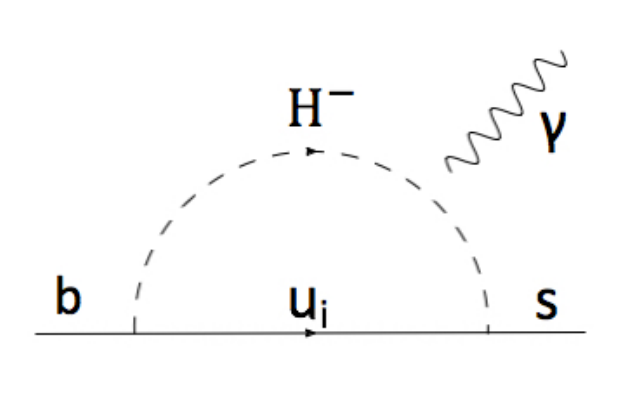}
    \caption{The diagram that contributes to the $b \to s \gamma$ process.}
    \label{diagram-bsgamma}
  \end{center}
\end{figure}

Next, we consider the rare decays of the mesons, such as $B \to X_s \gamma$.
The $b\to s$ transition is given by the $C_7$ operator, according to the diagram in Fig. \ref{diagram-bsgamma},
\beq
{\cal H}^{b \to s \gamma}_{eff} = - \frac{4G_F}{\sqrt{2}} V_{tb}V^*_{ts} \frac{e}{16 \pi^2} m_b C_7 F^{\mu \nu} ( \overline{s_L} \sigma_{\mu \nu} b_R) +h.c.,
\eeq
where $C_7$ in our model is evaluated at the one-loop level as follows:
\beq
C_7 =\frac{1}{4\sqrt{2}G_{\rm F} m_{H^+}^2 V_{tb}V_{ts}^*}\sum_i (V^\dagger \rho_u)^{si}(\rho_u^\dagger V)^{ib}
  \left [ \frac{2}{3} G^7_{ 1}(x_i)+ G^7_{2}(x_i)\right].
\eeq
$G^7_{ 1}(x)$ and $G^7_{2}(x)$ are defined as 
\begin{align}
  G^7_{ 1}(x)&=-\frac{2+3x-6x^2+x^3+6x\log x}{12(1-x)^4},\\
  G^7_{ 2}(x)&=-\frac{1-6x+3x^2+2x^3-6x^2\log x}{12(1-x)^4}.
\end{align}
The $b\rightarrow s\gamma$ has been experimentally measured and the result is consistent with the 
SM prediction \cite{Misiak:2015xwa}. Then, this process becomes a stringent bound on our model.
For instance, the size of $C_7$ at the bottom quark mass scale should be within the range,
$- 0.055 \le C_7 (m_b) \le 0.02$, according to the global fitting \cite{Cline:2015lqp}.

In Table \ref{table2}, we derive the upper bounds on the up-type Yukawa couplings using the value in Ref. \cite{Cline:2015lqp}. The charged Higgs mass, $m_{H^\pm}$, is fixed at $m_{H^\pm}=200$ GeV or $250$ GeV.
These results are consistent with the ones derived from the values in Refs. \cite{Blake:2016olu,Amhis:2016xyh}.
  \begin{table}[h]
  \begin{center}
    \begin{tabular}{|c|c|c|c|}
      \hline
$m_{H^\pm}$&$|\rho^{ct}_u|$&$|\rho^{tc}_u|$&$|\rho^{tt}_u|$\\
\hline
200 [GeV]&1.03&1.07&1.71\\
250 [GeV]&1.17&1.33&1.94\\
\hline 
    \end{tabular}
    \caption{The upper bounds from the global fitting: $-0.055\le\Delta C_7(m_b)\le0.02$.
    $m_{H^+}$ is fixed at $m_{H^\pm}=200$ GeV and $250$ GeV, respectively. }
    \label{table2}
  \end{center}
\end{table}


In addition, we could obtain the limits on the Yukawa couplings from the direct search
for the flavor-violating processes. 
In our model, the flavor-violating top quark decay is predicted as
\begin{align}
BR(\rm{t\rightarrow hc})=&\frac{|\rho^{tc}_u|^2+|\rho^{ct}_u|^2}{64\pi \Gamma_t}\cos^2 \theta_{\beta\alpha} \big(1-\frac{m_h^2}{m_t^2}\big)\nonumber\\
=&9.7\times10^{-4} ~\left (|\rho^{tc}_u|^2+|\rho^{ct}_u|^2 \right )\biggl(\frac{\cos \theta_{\beta\alpha}}{0.1}\biggl)^2,
\end{align}
where $\Gamma_t$ is defined as $\Gamma_t=1.41$GeV.
Based on the results in Refs \cite{Aad:2015pja,Khachatryan:2016atv,Aaboud:2017mfd},
we derive the following upper bound:
 \beq
|\cos \theta_{\beta\alpha} |\times\sqrt{|\rho_u^{tc}|^2+|\rho_u^{ct}|^2 }\le9.1\times10^{-2}.
\eeq
In our study, we survey the parameter region with ${\cal O}(1)$ $\rho_u^{tc}$ and/or $\rho_u^{ct}$.
In addition, $\rho_e^{\mu \tau}$ and $\rho_e^{\tau \mu}$ are large in some cases. 
As we discuss below, the flavor-violating Higgs decay, such as $h \to \mu \tau $, also significantly 
constraints $\cos \theta_{\beta\alpha}$. Then, we simply assume that $|\cos \theta_{\beta\alpha}|$
is at most ${\cal O}(10^{-3})$ and ignore the corrections that depend on $\cos \theta_{\beta\alpha}$.

\subsection{The experimental constraints on $\rho_e$ and $\rho_\nu$ }
\label{sec;rhoe}
In this section, we summarize the constraints on $\rho_e$ and $\rho_\nu$.
Interestingly, the texture of $\rho_e$ in Eq. (\ref{eq;rhoe}) can evade the strong experimental bounds
from the LFV processes.
On the other hand, the discrepancy of $(g-2)_{\mu}$ can be resolved by
the sizable ($\mu, \, \tau$) Yukawa couplings \cite{Omura:2015nja,Omura:2015xcg}.

Let us discuss the tree-level contributions to the physical observables, that are given by $\rho_e^{\mu \tau}$, $\rho_e^{\tau \mu}$ and $\rho_\nu$.
In the type-III 2HDM, the charged Higgs boson exchanging induces $\tau~\rightarrow~l \nu \overline{\nu}$ ($l= \mu, \, e$) at the tree level. We define the following observable:
\begin{align}
\biggl(\frac{g_\mu}{g_e}\biggl)^2\equiv\frac{BR(\tau~\rightarrow~\mu\nu\overline{\nu})/f(y_\mu)}{BR(\tau~\rightarrow~e\nu\overline{\nu})/f(y_e)},
\end{align}
where $y_{l}$~$\equiv m^2_l/m^2_\tau$ ($l=e,\, \mu, \,\tau$) are defined and $f(y)$ is a phase space function. 
This measurement has been experimentally given as $g_\mu / g_e = 1.0018\pm0.0014$ \cite{Patrignani:2016xqp}.
In our model, the extra contribution to the each branching ratio of $l_1 \to l_2 \nu \nu$ decay is proportional to 
\begin{equation}
|g^\nu_{l_1 l_2}|^2 \equiv \sum_{ij}  \left |  \left (\widetilde{ \rho_\nu} \right )^{l_2 i}\right |^2 \left | \left (\widetilde{ \rho_\nu} \right )^{l_1 j} \right |^2  , ~|g^e_{l_1 l_2}|^2 \equiv \sum_{ij}  \left | \left ( V_{\nu}^\dagger \, \rho_e \right )^{ i l_2 }\right |^2 \left | \left ( V_{\nu}^\dagger \, \rho_e \right )^{j l_1} \right |^2 ,
\end{equation}
where $\widetilde{ \rho_\nu}$ is defined as $\widetilde{ \rho_\nu} \equiv V_{\nu} \rho_\nu$.
Allowing the $2 \sigma$ deviation of $g_\mu / g_e$, we obtain the upper bounds on the Yukawa couplings at $m_{H^\pm} =$ 200(250) GeV as follows:
\beq
|g^\nu_{\mu \tau}| \leq 0.25 \, (0.4), ~|g^e_{\mu \tau}| \leq 0.25 \, (0.4).
\eeq

Next, we study the constraints from the michel parameter of the lepton decays.
As discussed above, the charged Higgs exchanging contributes to $l_1 \to l_2 \nu \bar{\nu}$ decays.
The constraints derived from the michel parameters are summarized in Ref. \cite{Patrignani:2016xqp}.
Following Ref. \cite{Patrignani:2016xqp}, we derive the bounds on $\rho_\nu$ and $\rho_e$ as 
\begin{equation}
\left | 0.76 \times g^\nu_{l_1 l_2} \left(\frac{200}{m_{H^\pm}}\right)^2 \right| \le c^\nu_{l_1 l_2}  
\end{equation}
and
\begin{equation}
\left | 0.76 \times g^e_{l_1 l_2} \left(\frac{200}{m_{H^\pm}}\right)^2 \right| \le c^e_{l_1 l_2}. 
\end{equation}
$c^\nu_{l_1 l_2} $ and $c^e_{l_1 l_2} $ are the upper bounds from $l_1 \to l_2 \nu \nu$, introduced in Ref. \cite{Patrignani:2016xqp}:
$( c^\nu_{\mu e},\, c^\nu_{\tau e},\, c^\nu_{\tau \mu} )= ( 0.55, \,2.01,\,2.01)$ and $( c^e_{\mu e},\, c^e_{\tau e},\, c^e_{\tau \mu} )= ( 0.035, \,0.70,\,0.72)$. Thus, we obtain the strong bounds on $g^\nu_{\mu e}$ and $g^e_{\mu e}$: $|g^\nu_{\mu e}|\leq 0.73 (1.13)$ and $|g^e_{\mu e}| \leq 0.046 (0.072)$ at $m_{H^\pm}=200 (250)$ GeV.
The other elements, on the other hand, can be ${\cal O}(1)$. 

Note that in our texture as Eq. (\ref{eq;rhoe}), $\rho_e^{\mu\mu}$ and $\rho_e^{ee}$ are assumed to be vanishing, so that 
the stringent constraints from the LFV decays of the charged leptons can be evaded.
In our setup with $\rho_e$ in Eq. (\ref{eq;rhoe}), the scalar mixing, $\cos \theta_{\beta \alpha}$, enhances the LFV $\tau$ decay, $\tau \to 3 \mu$, according to the neutral scalar exchanging. In order to avoid the current experimental bound, $Br(\tau \to 3 \mu)< 2.1 \times 10^{-8}$ \cite{Patrignani:2016xqp}, we obtain the bound as
\beq
|\cos \theta_{\beta \alpha}| \times  \overline{\rho_e^{\mu\tau}}  \lesssim 0.168 \times \left (1- \frac{(125\, {\rm GeV})^2}{m^2_H} \right )^{-1}, 
\eeq
where $ \overline{\rho_e^{\mu\tau}} \equiv \sqrt{|\rho_e^{\mu \tau}|^2+|\rho_e^{\tau \mu}|^2}$ is defined.  
Then we can conclude that the ($\mu$, $\tau$) elements of $\rho_e$ can be larger than ${\cal O}(0.1)$ when
 $|\cos \theta_{\beta\alpha}|$ is suppressed. 
In the case that $\rho_e^{\mu \mu}$ and $\rho_e^{ee}$ are sizable, the upper bounds on the parameters are
estimated as ${\cal O}(10^{-4})$ when the CP even scalar mass is ${\cal O}(100)$ GeV and $ \overline{\rho_e^{\mu\tau}}$ is ${\cal O}(1)$ \cite{ Omura:2015xcg}.

We can derive the constraint from the flavor-violating decay of 125 GeV neutral scalar.
In our model, the branching ratio of the decay to two fermions ($f_i$, $f_j$) is given by
\begin{align}
BR(h\rightarrow f_if_j)&=\frac{\Gamma(h\rightarrow f_i\bar{f_j})+\Gamma(h\rightarrow \bar{f_i}f_j)}{\Gamma_h} \nonumber \\
&=\frac{\cos^2 \theta_{\beta\alpha}\left (|\rho _f^{ij}|^2+|\rho _f^{ji}|^2 \right ) \, m_h}{16\pi\Gamma_h}, 
\end{align}
where $\Gamma_h$ is the total decay width of $h$ whose mass is around 125 GeV and fixed at $\Gamma_h$ = 4.1 MeV.
Following the upper bound on $BR(h\to \mu\tau)$ \cite{Khachatryan:2015kon,Aad:2016blu,CMS:2017onh},
we find the upper limit on the $\mu$-$\tau$ coupling at 2$\sigma$:
\beq
|\cos \theta_{\beta\alpha} | \times  \overline{\rho_e^{\mu\tau}} \le2.3\times10^{-3}.
\eeq
Thus, we obtain the strong bound on $\cos \theta_{\beta\alpha}$. As mentioned above, 
$|\cos \theta_{\beta\alpha}|$ is assumed to be at most ${\cal O}(10^{-3})$ and the contributions to
the flavor physics are ignored in our analysis.

We consider the one-loop contributions to the LFV process and the $Z$-boson decay.
The correction involving only $\rho_e$ is summarized in Ref. \cite{Omura:2015xcg}.
Assuming that the all elements of $\rho_e$ are vanishing, 
we derive the constraints on $\rho_\nu$ from the LFV processes.
The upper bounds from $l^\prime \to l \gamma$ are summarized in Table \ref{table4}.
$\Delta_{l l^\prime}$ is defined as
$\Delta_{l l^\prime}= \sum_j|(\widetilde{\rho_\nu})^{l j}(\widetilde{\rho_\nu})^{l^\prime j*}|$. 
As we see in Table \ref{table4}, $\Delta_{e \mu}$ is strongly constrained, while the other elements can be large. 

\begin{table}[h]
  \begin{center}
    \begin{tabular}{|c|c|c|c|}
      \hline
&$\Delta_{\mu \tau}$ &$\Delta_{e \tau}$ & $\Delta_{e \mu }$\\
\hline
$m_{H^\pm}$=200 [GeV]&0.135& 0.116&0.173$\times10^{-3}$ \\
\hline
$m_{H^\pm}$=250 [GeV]&0.211&0.181& 0.275$\times10^{-3}$\\
\hline
 \end{tabular}
    \caption{The upper bounds on $\widetilde{\rho_\nu}=V_{\nu} \rho_\nu$ at 90$\%$ CL in the cases with 
    $m_{H^\pm}=200$ GeV and $250$ GeV. 
    $\Delta_{l l^\prime}= \sum_j|(\widetilde{\rho_\nu})^{l j}(\widetilde{\rho_\nu})^{l^\prime j*}|$ is defined.}
    \label{table4}
  \end{center}
\end{table}


In addition, the decay of the $Z$ boson may be largely deviated from the SM prediction, according to
the extra scalars, at the one-loop level.
In our work, we consider the case that either $\rho_e$ or $\rho_\nu$ is sizable.
Then, the contribution to the $Z$ boson decay through the penguin diagrams is suppressed.
We have calculated the deviation of $BR(Z\to \nu\bar\nu)$, but it is not so large.
We find that the upper bounds on $\left|\overline{\rho_e^{\mu \tau}}\right|$ and $|\overline{\rho_\nu}|$ 
can reach ${\cal O}(1)$, even if the deviation of $BR(Z\to \nu\bar\nu)$ is required to be within 
2$\sigma$.

Note that $\rho_\nu$ would be strongly constrained by the cosmological observation, depending on the
mass spectrum of the right-handed neutrino. We comment on the bound in Sec. \ref{sec;caseC}.


\section{The (semi)leptonic $B$ decays}
\label{sec;flavor2}
Based on the studies in Sec. \ref{sec;flavor},
we investigate the impact of our Type-III 2HDM on the (semi)leptonic $B$-meson decays.
As discussed in Refs. \cite{Hu:2016gpe,Ko:2017lzd,Iguro:2017ysu},
the 2HDMs potentially have a great impact on $B \to D^{(*)} l \nu$ and 
$B \to K^{(*)} ll$ processes $(l=e, \, \mu, \, \tau)$, where the discrepancies between the
experimental results and the SM predictions are reported.
In particular, the global analyses on $B \to K^{(*)} ll$ suggest
that $C_9$ and $C_{10}$ operators may be deviated from the SM values.
Besides, the flavor universality of $B \to K^{(*)} ll$ is also inconsistent 
with the SM prediction in the experimental results.
In our model, $\rho_\nu$ can contribute to 
the $C_9$ and $C_{10}$ operators, flavor-dependently.
Thus, it becomes very important to find how well 
the tension can be relaxed, taking into account the $\rho_\nu$ contribution.

\subsection{The bounds from the  $B \to l \nu$ decays}
\label{Btolnu}
First, we discuss the leptonic decays of the $B$ meson: $B \to l \nu$.
In our model, the charged Higgs exchanging contributes to the $B$ meson decays as
\beq
\label{leptonicBdecay}
{\cal H}^l_{B_q}= -\frac{\rho_e^{l^\prime l}  \rho_u^{tq} }{m^2_{H^{\pm}}} (V^*_{\nu})_{l^\prime j} V_{tb} (\overline{\nu^j_L} l_R) (\overline{b_L} q_R) -\frac{ (\widetilde{\rho_\nu})^{l j*}  \rho_u^{tq} }{m^2_{H^{\pm}}}  V_{tb} (\overline{\nu^j_R} l_L) (\overline{b_L} q_R).
\eeq
The flavor of the neutrino in the final state can not be distinguished, so that let us define the parameters,
\beq
|\kappa^e_{l q}|^2 \equiv \sum_j \left | \rho_e^{j l}  \rho_u^{tq}  \right |^2, ~|\kappa^\nu_{l q}|^2 \equiv \sum_j \left | (\widetilde{\rho_\nu})^{l j*}  \rho_u^{tq} \right |^2,
\eeq
and discuss the constraints on those products.

In our setup, the ($t$, $c$)-elements of $\rho_u$ are sizable, so that
the leptonic decay of $B_c$ is deviated from the SM prediction.
The leptonic decay has not been measured by any experiments, but
we can derive the constraint from the total decay width of $B_c$ \cite{Alonso:2016oyd} 
and the measurement at the LEP experiment \cite{Akeroyd:2017mhr}.
Adopting the severe constraint, $BR(B_c \to \tau \nu) \leq 10$ \% \cite{Akeroyd:2017mhr},
we obtain the upper bounds on the lepton Yukawa couplings as follows:
\beq
\label{Bc}
|\kappa^{e,\nu}_{\tau c}| \times \left ( \frac{200\, {\rm GeV}}{m_{H^\pm}} \right )^2 \leq 0.025 .
\eeq

In our assumption, the ($t$, $u$) elements are less than ${\cal O}(0.01)$.
Even in such a case, the sizable $\rho_e^{\mu \tau, \tau \mu}$ and
$\rho_\nu$ may largely contribute the $B_u$ decays.
The contributions to $B_u \to l\bar{\nu}$ are linear to $|\kappa^e_{l u}|^2$ and $|\kappa^\nu_{l u}|^2$.
These products at $m_{H^\pm}=200(250)$ GeV are constrained by the leptonic $B_u$ decays as
\begin{eqnarray}
|\kappa^{e, \nu}_{\mu u}|&\leq&  0.99\times10^{-4}  ~(1.55\times10^{-4} ),  \\
|\kappa^{e, \nu}_{\tau u}|&\leq&  1.18\times10^{-3}~(1.84\times10^{-3}).  
\end{eqnarray}

We could also derive the bound from $B_s \to \mu \mu$.
The error of the experimental measurement is still so large that 
it is difficult to draw a stringent bound on our model.
The branching ratio of this rare decay, however, relates to the semi-leptonic $B$ decay, $B \to K^{(*)} \mu \mu$,
so that we give a discussion about this process below.


\subsection{$B \to D^{(*)} l \nu$ ($l=e, \, \mu, \, \tau$)}
\label{BtoDlnu}
We investigate the constraints from the semileptonic $B$ decay; e.g.,
$B \to D^{(*)} l \nu$ ($l=e, \, \mu, \, \tau)$. 
There is a discrepancy in $B \to D^{(*)} \tau \nu$, although
$B \to D^{(*)} e \nu$ and $B \to D^{(*)} \mu \nu$ are consistent with the SM predictions.
In our model, the charged Higgs exchanging flavor-dependently contributes to
these processes via $\rho_{e}$ and $\rho_{u}$ couplings, as shown in Eq. (\ref{leptonicBdecay}).
\begin{figure}[H]
\begin{center}
\includegraphics[width=4.5cm]{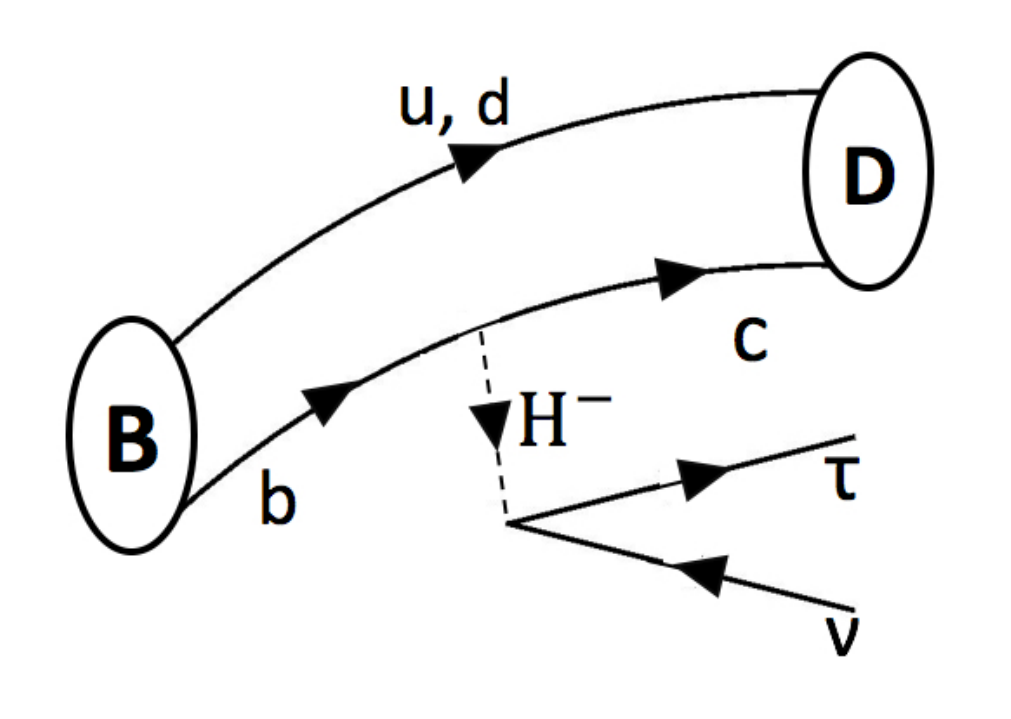}
\end{center}
\caption{ Diagram that contributes to the $B \to D \tau \nu$.}
\label{RD}
\end{figure}
Then, the discrepancy of $B \to D^{(*)} \tau \nu$ could be ameliorated by the contribution of the diagram in Fig. \ref{RD} \cite{Iguro:2017ysu}, although the flavor universality of $B \to D^{(*)} l \nu$ ($l=e, \, \mu)$
may constrain our setup strongly. We define the observables to measure the universality as follow:
\beq
R(D^{(*)})_{e\mu}= \frac{BR(B\to D^{(*)}e\bar{\nu})}{BR(B\to D^{(*)}\mu\bar{\nu})}.
\eeq
The deviations should not exceed a few percent: $R(D^{*})_{e\mu}=1.04 \pm 0.05 \pm 0.01$ \cite{Abdesselam:2017kjf}.
Fixing the charged Higgs mass at $m_{H^\pm}=200(250)
$ GeV, we derive the upper bounds on $\kappa^{e, \nu}_{\mu c}$ .
In the Table \ref{table8}, the upper bounds on $\kappa^{e, \nu}_{\mu c}$ with $m_{H^\pm}= 200$ GeV and 250 GeV are summarized. The calculation is based on Ref. \cite{Iguro:2017ysu}.
Note that, roughly speaking, only $BR(B\to D^{(*)}\mu\bar{\nu})$ is always enhanced, so the only lower limit on $R(D^{(*)})_{e\mu}$ is shown in Table \ref{table8}.
We impose the bounds as $R(D^{(*)})_{e\mu} > 0.95$ and $R(D^{(*)})_{e\mu} > 0.98$ \cite{Abdesselam:2017kjf}.

\begin{table}[h]
  \begin{center}
    \begin{tabular}{|c|c|c|c|}
      \hline
&$m_{H^\pm}$&$R(D^{(*)})_{e\mu}=0.95$&$R(D^{(*)})_{e\mu}=0.98$\\
\hline
$D^*$&200 [GeV]&$5.16\times10^{-2}$&$0.34\times10^{-1}$\\
\hline
$D^*$&250 [GeV]&$8.06\times10^{-2}$&$5.32\times10^{-2}$\\
\hline
$D$&200 [GeV]&$1.08\times10^{-2}$&$0.70\times10^{-2}$\\
\hline
$D$&250 [GeV]&$1.68\times10^{-2}$&$1.08\times10^{-2}$\\
\hline
 \end{tabular} 
    \caption{The upper bounds on $|\kappa^{e, \nu}_{\mu c}|$ from the lepton universality of $B \to D^{(*)} l \nu$ ($l=e, \, \mu$).
    We impose the upper bounds on $R(D^{(*)})_{e\mu}$ as $R(D^{(*)})_{e\mu} > 0.95$ and $R(D^{(*)})_{e\mu} > 0.98$ \cite{Abdesselam:2017kjf}. The process, $B \to D^{*} ( D) l \nu$, is labeled as $D^*$ ($D$) on the first column.  }
    \label{table8}
  \end{center}
\end{table}

The semileptonic $B$ decay associated with $\tau$ lepton in the final state is also deviated from the SM prediction,
in our model. 
In Ref. \cite{Iguro:2017ysu}, $R(D^{(*)})$ are well studied in the Type-III 2HDM with only $\rho_e$, $\rho_d$ and $\rho_u$, and we find that at least $R(D)$ can be enhanced so much that it is consistent with the experimental result.
The lowest value to achieve the world average of $R(D)$ ($R(D)$=0.407$\pm$0.046) and $R(D^*)$ ($R(D^*)$=0.304$\pm$0.015) at the 1$\sigma$ level \cite{Amhis:2016xyh} is
\begin{align}
|\kappa^{e,\nu}_{\tau c}| &\geq 2.12\times10^{-2} ~{\rm{for}}~R(D),\\
|\kappa^{e,\nu}_{\tau c}| &\geq 2.89\times10^{-1} ~{\rm{for}}~R(D^*),
\end{align}
when the charged Higgs mass is fixed at $m_{H^\pm}=200 $ GeV.
Note that our SM prediction is $R(D)=0.299$ and $R(D^*)=0.253$ with BR($B_c \to \tau \nu)=2.2\%$ in our parameter set \cite{Tanaka2010se}.
This lowest value for $R(D)$ is very close to the upper bound from $B_c \to \tau \nu$ in Eq. (\ref{Bc}).
We can find that the value required by $R(D^*)$ is totally excluded by the $B_c$ decay.
Besides, the lepton universality of this semileptonic decay provides the stringent bounds on $\kappa^{e, \nu}_{\mu c}$,
as shown in Table \ref{table8}. Thus, we concluded that either $|\kappa^{e}_{\tau c}|$ or $|\kappa^{\nu}_{\tau c}|$ should be ${\cal O}(1) \times 10^{-2}$ to achieve the discrepancy of $R(D)$ without any conflict with the other observables concerned with the $B$ decay. Otherwise, the anomaly of $R(D)$ cannot be resolved in our model.

\subsection{$B \to K^{(*)} ll$}

Finally, we consider $B \to K^{(*)} ll$ in our model.
In the so-called aligned 2HDM, this process has been discussed in Ref. \cite{Hu:2016gpe}.
The Type-III 2HDM case with only $\rho_e$ has also been shortly studied in Ref. \cite{Iguro:2017ysu}.
In our study, we include the box diagrams induced by $\rho_e$ and $\rho_\nu$
and take into account the consistency with the explanations of $(g-2)_{\mu}$ and $R(D)$,
that has not been done before. 

In the $B \to K^{(*)} ll$ processes, there are several interesting observables where the
discrepancies between the SM predictions and the experimental results are reported by
the LHCb collaboration. One is $P^\prime_5$ that is concerned with
the angular distribution of the $B \to K^{*} \mu \mu$ process \cite{Aaij:2013qta,Aaij:2015oid}, and
another is $R(K^{*})$ \cite{LHCbnew} and $R(K)$ \cite{Aaij:2014ora} that measure the lepton universalities
of $B \to K^{*} \mu\mu/ee$ and $B \to K \mu\mu/ee$, respectively.
The observables are governed by $C^l_9$ and $C^l_{10}$ operators defined as
\begin{equation}
{\cal H}^l_{B_s}= -g_{\rm SM}
\left  \{ C_9^{l}(\overline{s_L} \gamma_\mu b_L)
(\overline{l}\gamma^\mu l )
+C_{10}^{l}(\overline{s_L} \gamma_\mu b_L)
(\overline{l}\gamma^\mu \gamma_5 l) +h.c. \right \},
\end{equation}
where $g_{\rm SM}$ is the factor from the SM contribution:
\begin{equation}
g_{\rm SM} =\frac{4 G_F}{\sqrt{2}} V_{tb} V^*_{ts} \frac{e^2}{16 \pi^2}.
\end{equation}
In our model, the Wilson coefficients $C_{9}^{l}$ and $C_{10}^{l}$ 
consist of the SM and the new physics contributions
as $C_{9}^{l}=(C_{9})_{{\rm SM}} +\Delta C_{9}^{l}$ and $C_{10}^{l}=(C_{10})_{{\rm SM}} +\Delta C_{10}^{l}$.
$\Delta C^l_9$ and $\Delta C_{10}^{l}$ are given by
\begin{align}
    \Delta C_{9(l)}& =\frac{-1}{2\sqrt{2} G_{\rm F}m_{H^+}^2 V_{tb}V_{ts}^*}
     \sum_i (V^\dagger \rho_u)^{si}(\rho_u^\dagger V)^{ib}
    \left[ \frac{2}{3} G_{\gamma 1}(x_i)+ G_{\gamma 2}(x_i)\right]\nonumber \\
    &+\frac{1}{4\pi \alpha V_{tb} V_{ts}^*} \left(- \frac{1}{2}+2 s_W^2 \right )\sum_i (V^\dagger \rho_u)^{si}(\rho_u^\dagger V)^{ib} G_Z(x_i),\\
    \Delta C_{10(l)} &=\frac{1}{4\pi \alpha V_{tb} V_{ts}^*} \, \frac{1}{2} \, \sum_i (V^\dagger \rho_u)^{si}(\rho_u^\dagger V)^{ib} G_Z(x_i),
    \label{C10_general}
\end{align}
where $s_W$ corresponds to the Weinberg angle and the functions are defined as
\begin{align}
G_{\gamma 1}(x)&=-\frac{16-45x+36x^2-7x^3+6(2-3x)\log x}{36 (1-x)^4},\\
G_{\gamma 2}(x)&=-\frac{2-9x+18x^2-11x^3+6x^3\log x}{36 (1-x)^4},\\
G_Z(x) &=\frac{x(1-x+\log x)}{2(1-x)^2}.
\end{align}
We note that the SM predictions are flavor universal and the size of the each coefficient at the bottom mass scale is estimated as $(C_{9})_{{\rm SM}} \approx 4$ and $(C_{10})_{{\rm SM}} \approx -4$, respectively.

The excesses in both $P^\prime_5$ and $R(K^{(*)})$ require destructive interferences with the SM predictions;
for instance, the $1 \sigma$ region of $|\Delta C_{9}^{\mu}|$ suggested by the global analysis is $-0.81 \leq \Delta C_{9}^{\mu} \leq -0.48$ ($1 \sigma$) and $-1.00 \leq \Delta C_{9}^{\mu} \leq -0.32$ ($2 \sigma$),
assuming $\Delta C_{9}^{\mu}=- \Delta C_{10}^{\mu}$ \cite{Altmannshofer:2017yso}.
There are a lot of works on the global fitting \cite{Altmannshofer:2017yso,Descotes-Genon:2013wba,Hiller:2014yaa,Altmannshofer:2015sma,Descotes-Genon:2015uva,Hurth:2016fbr,DAmico:2017mtc,Ciuchini:2017mik,Bardhan:2017xcc}.
The results are consistent with each other and the excesses require large contributions to
the muon couplings: $(\Delta C_9^{l})/(C_9)_{{\rm SM}} \simeq -0.2$ and $(\Delta C_{10}^{l})/(C_{10})_{{\rm SM}} \simeq 0.2$. We note that $\Delta C_{10}^{l}$ need not be large, while such a large $\Delta C_9^{l}$ is favored.
In fact, the scenario with vanishing $\Delta C_{10}^{l}$ can fit the experimental results at the $2 \sigma$ level \cite{DAmico:2017mtc}.

It is important that these observables have different characteristics: $R(K^{(*)})$ requires the violation of the flavor universality, but $P^\prime_5$ does not need the violation. 
In our study, we concentrate on the three cases:
\begin{enumerate}
\renewcommand{\labelenumi}{(\Alph{enumi})}
\item $\rho_e^{ij}=0$ and $\rho_\nu^{ij}=0$,
\item $\rho_e^{\mu \tau} \neq 0$, $\rho_e^{\tau \mu} \neq 0$ and $\rho_\nu^{ij}=0$,
\item  $\rho_e^{ij} = 0$ and $(\widetilde{\rho_\nu})^{\mu j} \neq 0$.
\end{enumerate}

In the case (A), the extra scalars do not couple to leptons, so that
we can not expect the violation of the lepton universality. 
$P^\prime_5$ in this framework has been studied in Refs. \cite{Hu:2016gpe,Iguro:2017ysu},
and we find the sizable $\rho_u^{tc}$, $\rho_u^{ct}$ and $\rho_u^{tt}$ lead large $\Delta C_{9}$ and $\Delta C_{10}$.

In the case (B), $\rho_e^{\mu \tau}$ and $\rho_e^{\tau \mu}$ are only non-vanishing.
In such a case, we can expect that the discrepancy of $(g-2)_{\mu}$ is explained by
the one-loop correction involving the neutral scalars \cite{Omura:2015nja,Omura:2015xcg}.  
Besides, the violation of the lepton universality in $B \to K^{(*)} ll$ would be realized,
if $\rho_u^{tc}$, $\rho_u^{ct}$ and $\rho_u^{tt}$ are sizable.

In the case (C), we assume that $(\widetilde{\rho_\nu})^{\mu j}$ is only sizable.
In this case, the box diagram involving the charged Higgs leads the destructive interference
with the SM prediction in $C^\mu_9$ and $C^\mu_{10}$, so that the anomaly of
$R(K^{(*)})$ may be resolved. 

Below, we discuss the induced $C_9$, $C_{10}$ and the relevant constraints in the each case.
We do not consider the case that both $(\widetilde{\rho_\nu})^{\mu j}$ and $\rho_e^{\mu \tau,\tau \mu}$
are sizable, in order to avoid the left-right mixing couplings of leptons induced by the one-loop diagrams involving 
the extra scalars. 

\subsubsection{Case (A): $\rho_e^{ij}=0$ and $\rho_\nu^{ij}=0$}
\begin{figure}[H]
\begin{center}
\includegraphics[width=4.5cm]{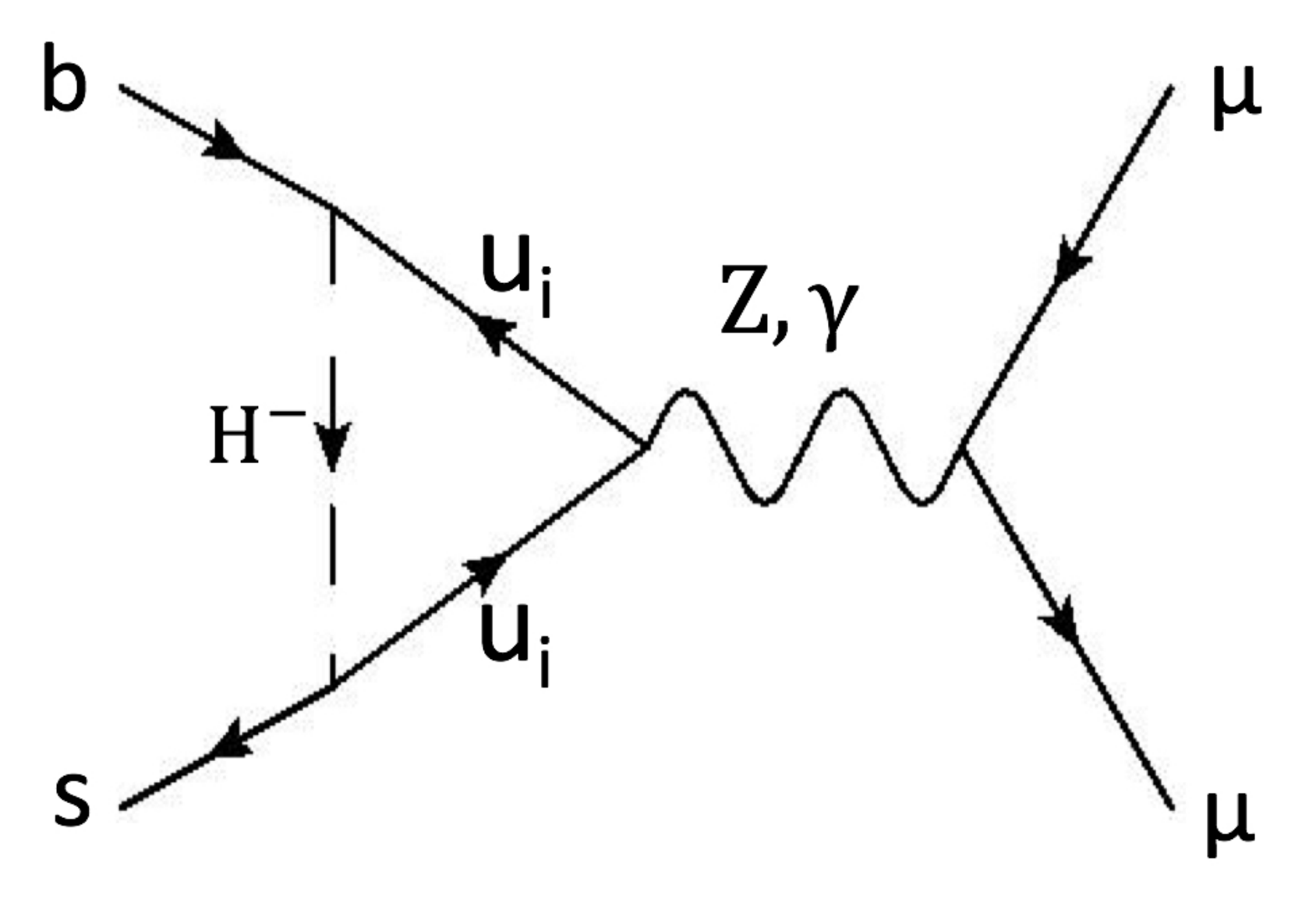}
\end{center}
\caption{ Diagram that contributes to the $B \to K \mu\mu$ in all cases.}
\label{BtoKmumu-penguin}
\end{figure}

\begin{figure}[t]
  \begin{center}
    \includegraphics[width=0.4\textwidth]{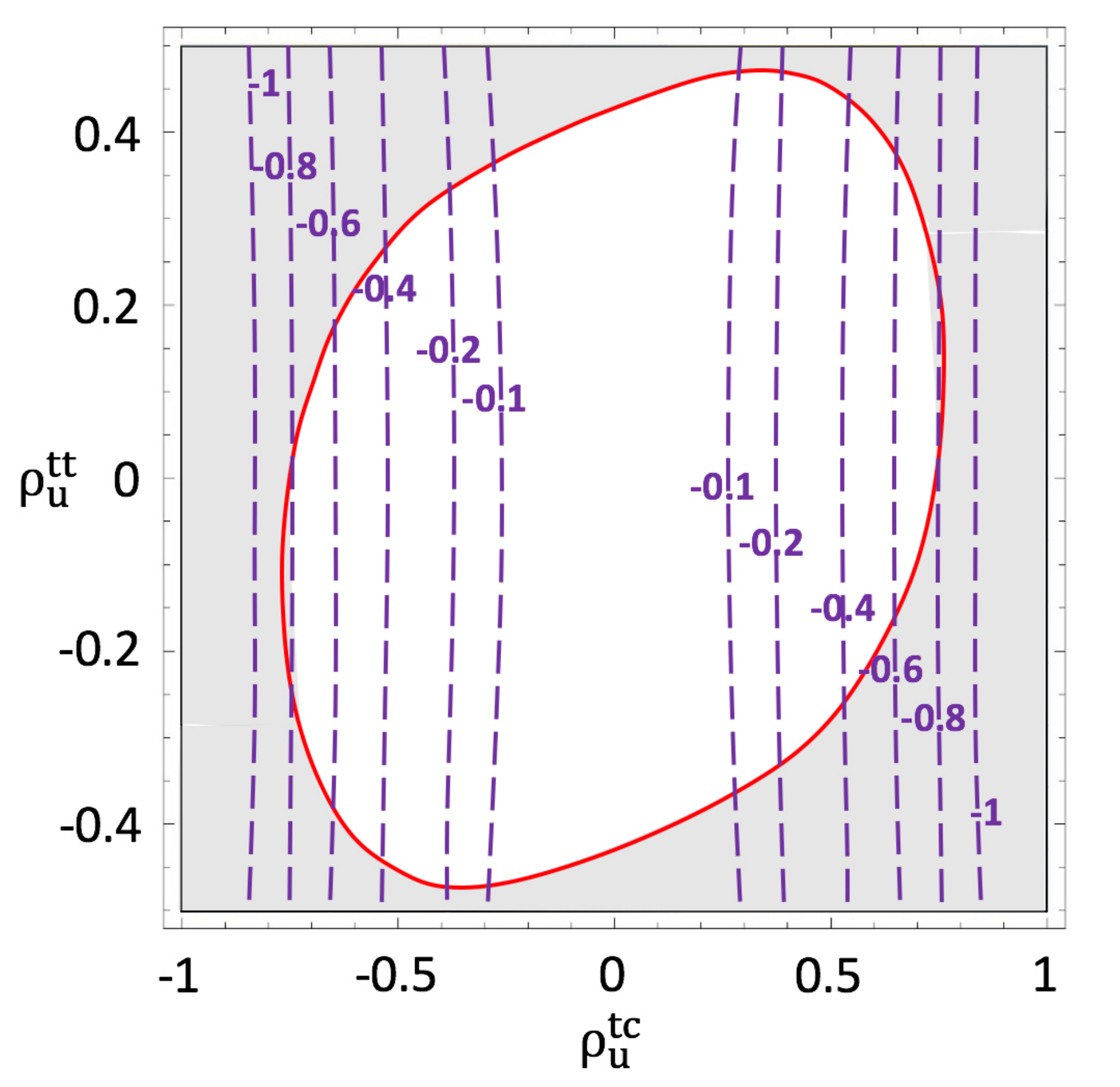}
    \includegraphics[width=0.4\textwidth]{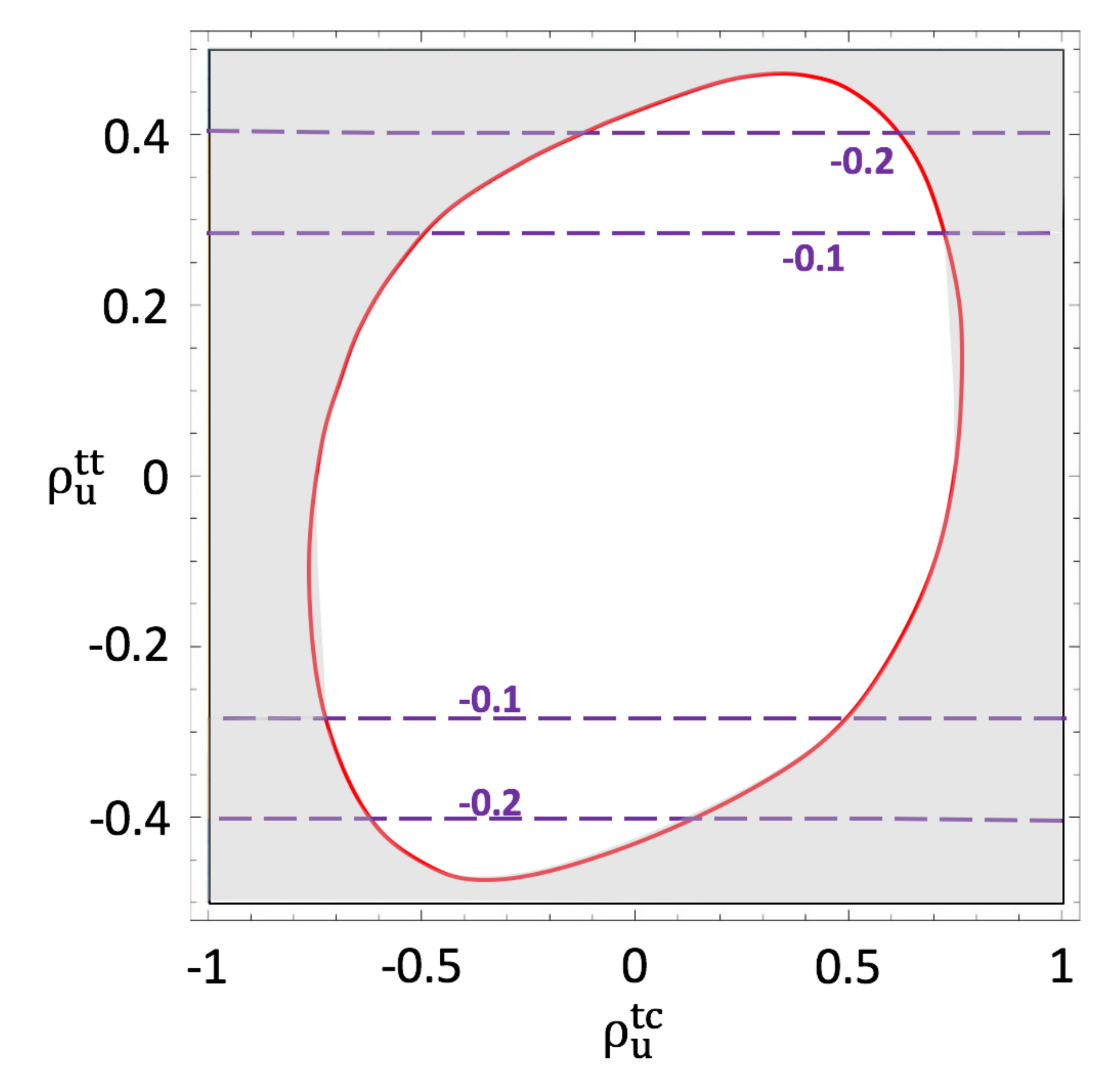}
    \includegraphics[width=0.4\textwidth]{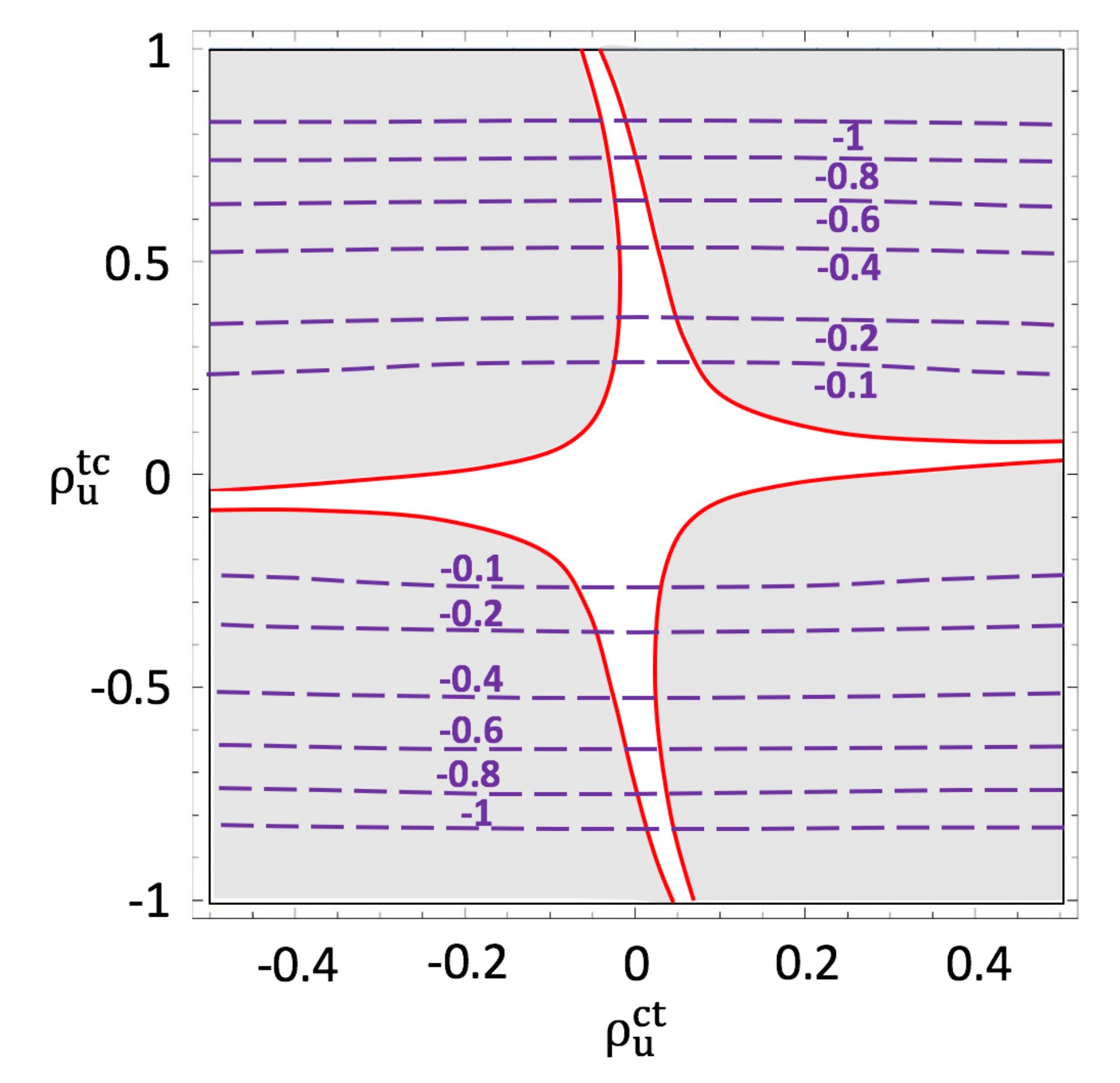}
    \includegraphics[width=0.4\textwidth]{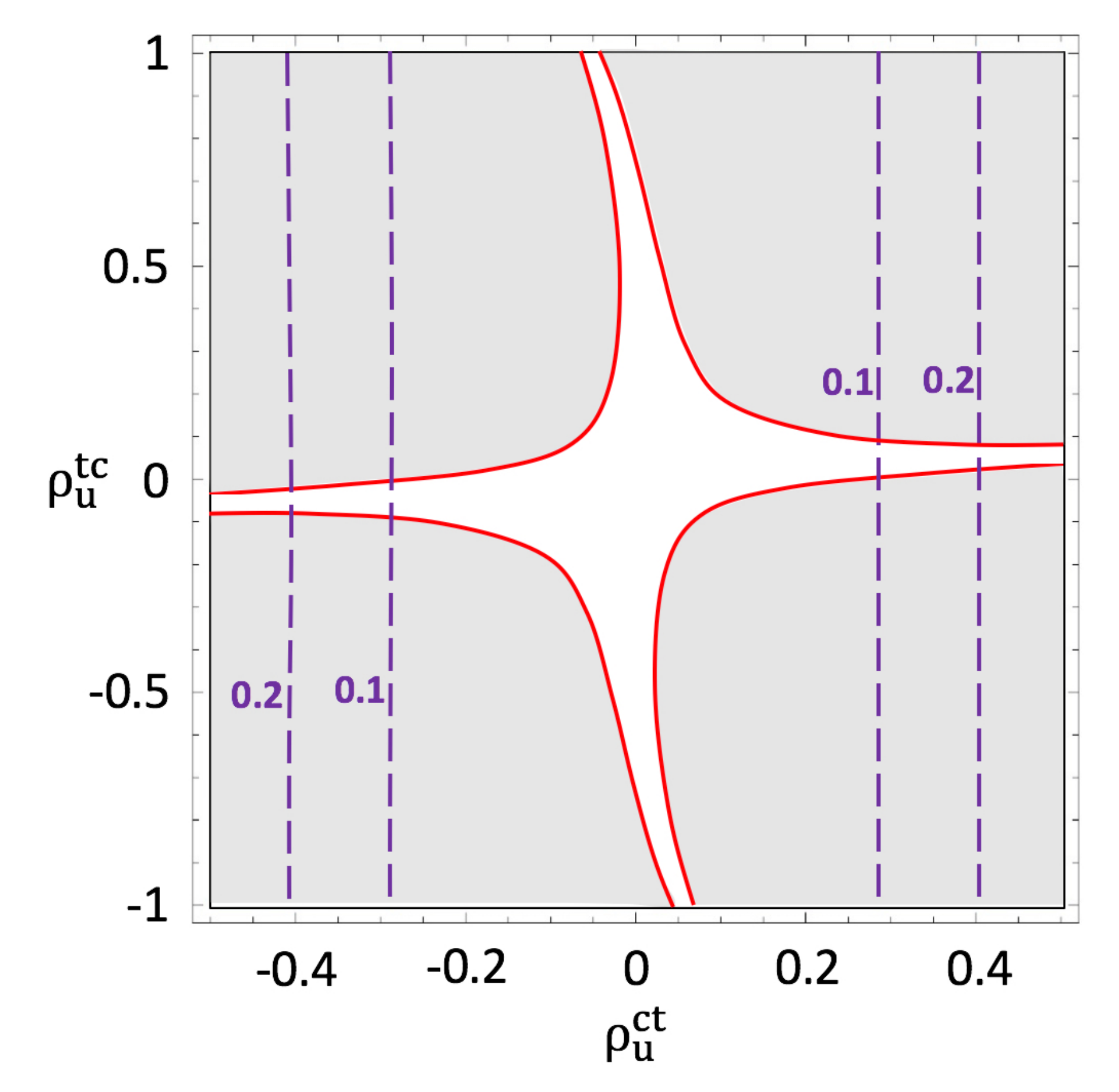}
    \caption{$\rho_u^{tt}$ vs. $\rho_u^{ct}$ (upper) and $\rho_u^{tc}$ vs. $\rho_u^{ct}$ (lower) in the case (A) with $m_{H^\pm}=200$ GeV. The gray region is excluded by the $B_s-\overline{B_s}$ mixing and the red lines correspond to the borders. The dashed purple lines denote the predictions of $\Delta C_9$ (left) and $\Delta C_{10}$ (right).}
    \label{scenario1}
  \end{center}
\end{figure}

In the case (A), the violation of the lepton universality can not be expected, but
large $\Delta C_9$ and $\Delta C_{10}$ may be induced by the loop diagrams involving the scalars.
In our setup, the main contributions to the operators are given by the couplings, $\rho_u^{tc}$, $\rho_u^{ct}$,
and $\rho_u^{tt}$. Then, the charged Higgs plays a crucial role in $\Delta C_9$ and $\Delta C_{10}$. 
The dominant contribution is given by the penguin diagram in Fig. \ref{BtoKmumu-penguin}.
We note that this type diagram is allowed in all cases.
Setting the charged Higgs mass at $m_{H^{\pm}}=200$ GeV, we draw the predicted $\Delta C_9$ and $\Delta C_{10}$ in Fig. \ref{scenario1}. The relevant constraints are shown in those plots. The gray region is excluded by the $B_s-\overline{B_s}$ mixing in Fig. \ref{scenario1}. Note that the constraint from the $b \to s \gamma$ process is out of the figures. The dashed purple lines denote the predictions of $\Delta C_9$ and $\Delta C_{10}$ on the left and right panels.
The size of the deviation is denoted on the each line. In the figures on the upper (lower) line, $\rho_u^{ct}$ ($\rho_u^{tt}$)
is assumed to be vanishing. We see that $\rho_u^{tt}$ does not help the enhancement of $\Delta C_9$, but either $\rho_u^{tc}$ or $\rho_u^{ct}$ can achieve $\Delta C_9 \approx -1$, that can explain the $P^\prime_5$ excess within $1 \sigma$ level. We note that $\rho_u^{tc}$ is not sensitive to $\Delta C_{10}$.

Let us comment on the contribution to the $B_s \to \mu \mu $ process.
The positive (negative) $\Delta C^{\mu}_{10}$ coefficient suppresses (enhances)
the branching ratio, compared to the SM prediction. The experimental result 
still has a large uncertainty, and the central value is below the SM prediction \cite{Aaij:2017vad}.
Thus, the positive $\Delta C^{\mu}_{10}$ is, in effect, favored, taking into account 
the $B_s \to \mu \mu $ process as well \cite{DAmico:2017mtc}. If we chose the parameter to predict $\Delta C^{\mu}_{10} \simeq 0.1$, the suppression is about 2.4 \%.

\subsubsection{Case (B): $\rho_e^{\mu \tau} \neq 0$, $\rho_e^{\tau \mu} \neq 0$ and $\rho_\nu^{ij}=0$ }

In the case (B), we consider the scenario that both $\rho_e^{\mu \tau}$ and $\rho_e^{ \tau \mu}$
are sizable, motivated by the $(g-2)_\mu$ anomaly.
Note that the mass difference between $H$ and $A$ is also required to explain the excess \cite{Omura:2015nja,Omura:2015xcg}. As discussed in Sec. \ref{Btolnu} and Sec. \ref{BtoDlnu}, $\rho_u^{tc}$ leads the conflict with $B \to D^{(*)} l \nu$ processes, if $\rho_e^{\mu \tau,\tau \mu}$ are sizable. 
The deviation of $(g-2)_\mu$, denoted by $\delta\alpha_\mu$, is evaluated at the one-loop level as
\begin{equation}
\delta\alpha_\mu=2.61\biggl(\frac{\rho_e^{\tau\mu}\rho_e^{\mu\tau}}{-0.034}\biggl)\times10^{-9},
\end{equation}
when $(m_A, \, m_H)$ is fixed at $(m_A, \, m_H)=(200 \, {\rm GeV}, \, 250 \, {\rm GeV})$.
The value experimentally required \cite{Hagiwara:2011af}\footnote{See also Refs.\cite{Davier:2016iru,Jegerlehner:2017lbd,Hagiwara2017zod} for a recent development.} is $\delta\alpha_\mu=(2.61\pm0.8)\times10^{-9}$  
, so that $\rho_e^{\tau\mu}\rho_e^{\mu\tau}$ should be about $0.03$
to explain the discrepancy at the $1 \sigma$ level.
\begin{figure}[t]
  \begin{center}
    \includegraphics[width=0.4\textwidth]{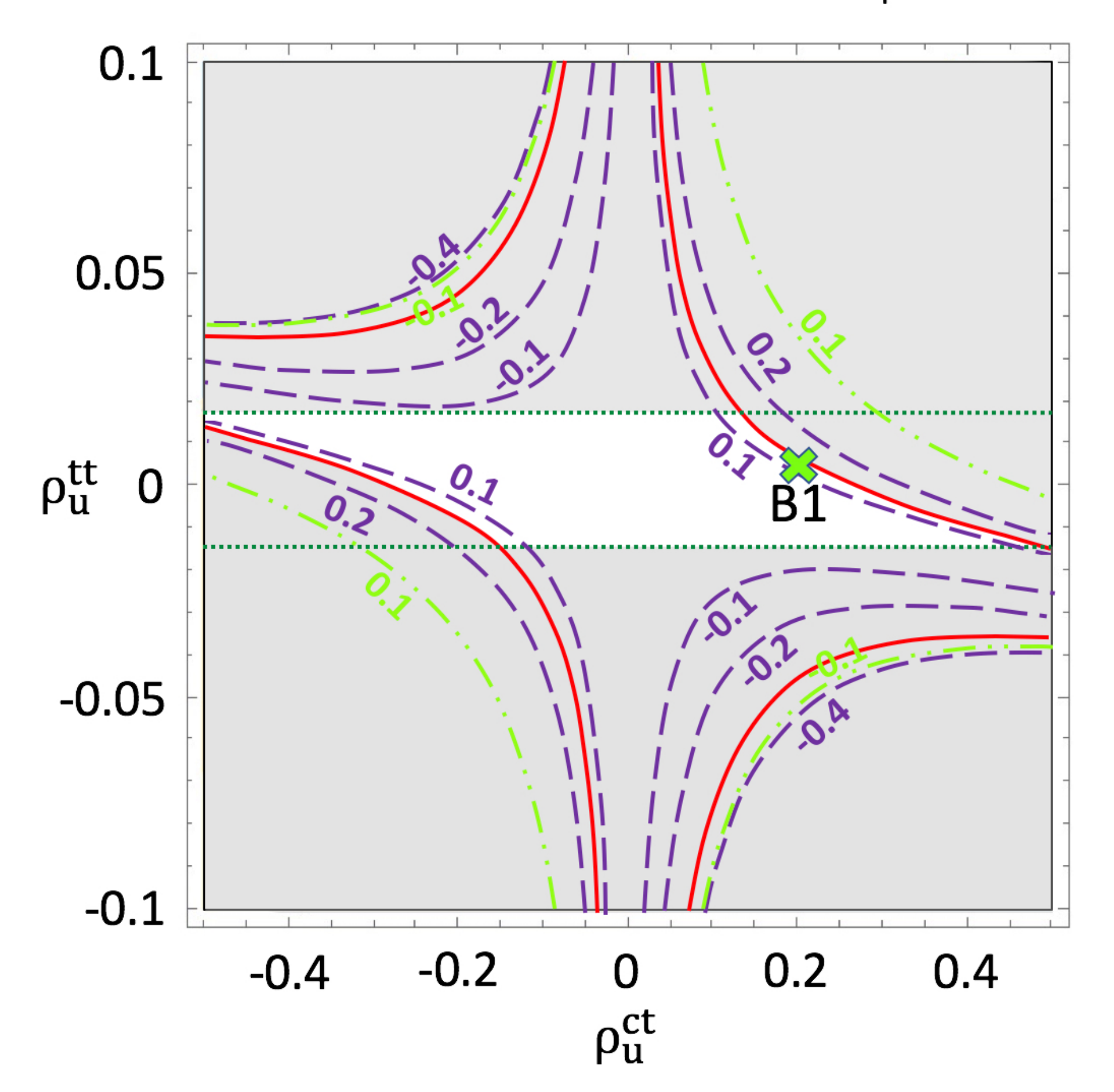}
 \includegraphics[width=0.4\textwidth]{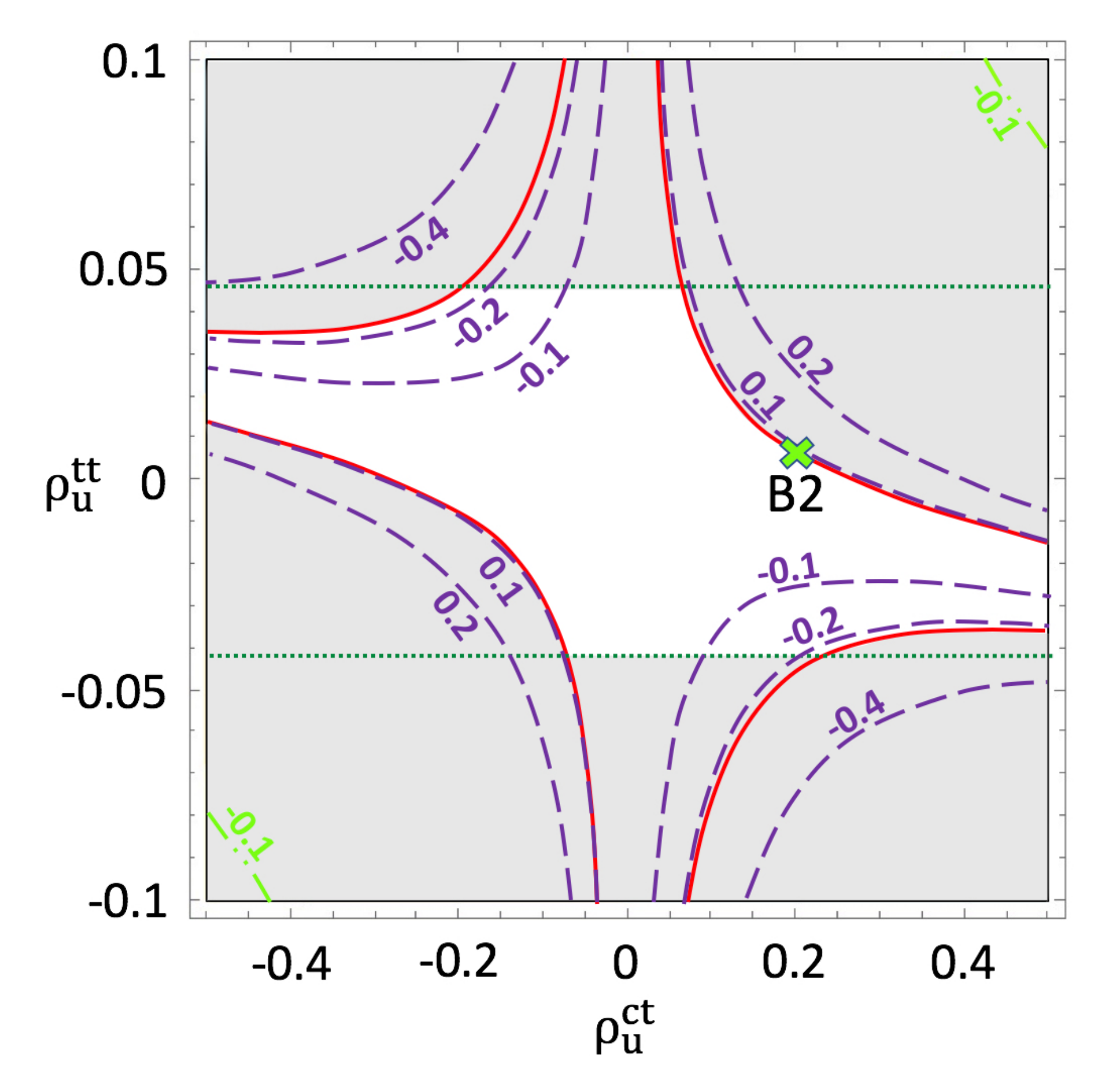}
        \caption{$\rho_u^{tt}$ vs. $\rho_u^{ct}$ in the case (B) with $\rho_e^{\tau\mu}=1(\rm{left}), \, 0.1(\rm{right})$ and $(m_A, \, m_H, \, m_{H_\pm})=(200 \, {\rm GeV}, \, 250 \, {\rm GeV}, \, 200 \, {\rm GeV})$. 
        $\rho_e^{\mu \tau}$ is fixed at  $\rho_e^{\mu \tau}=-0.034, \, -0.34$ that correspond to 
   $\delta\alpha_\mu=(2.61)\times10^{-9}$. The gray region is excluded by the $B_s-\overline{B_s}$ mixing (red lines) and $\tau \to  \mu\gamma$ process (dotted green lines). The dashed green lines and dashed purple lines denote the predictions of $\Delta C_9$ and $\Delta C_{10}$ for the each case. The size of the deviation is shown on the each line.}
    \label{scenario3}
  \end{center}
\end{figure}

\begin{figure}[h]
\begin{center}
\includegraphics[width=4.5cm]{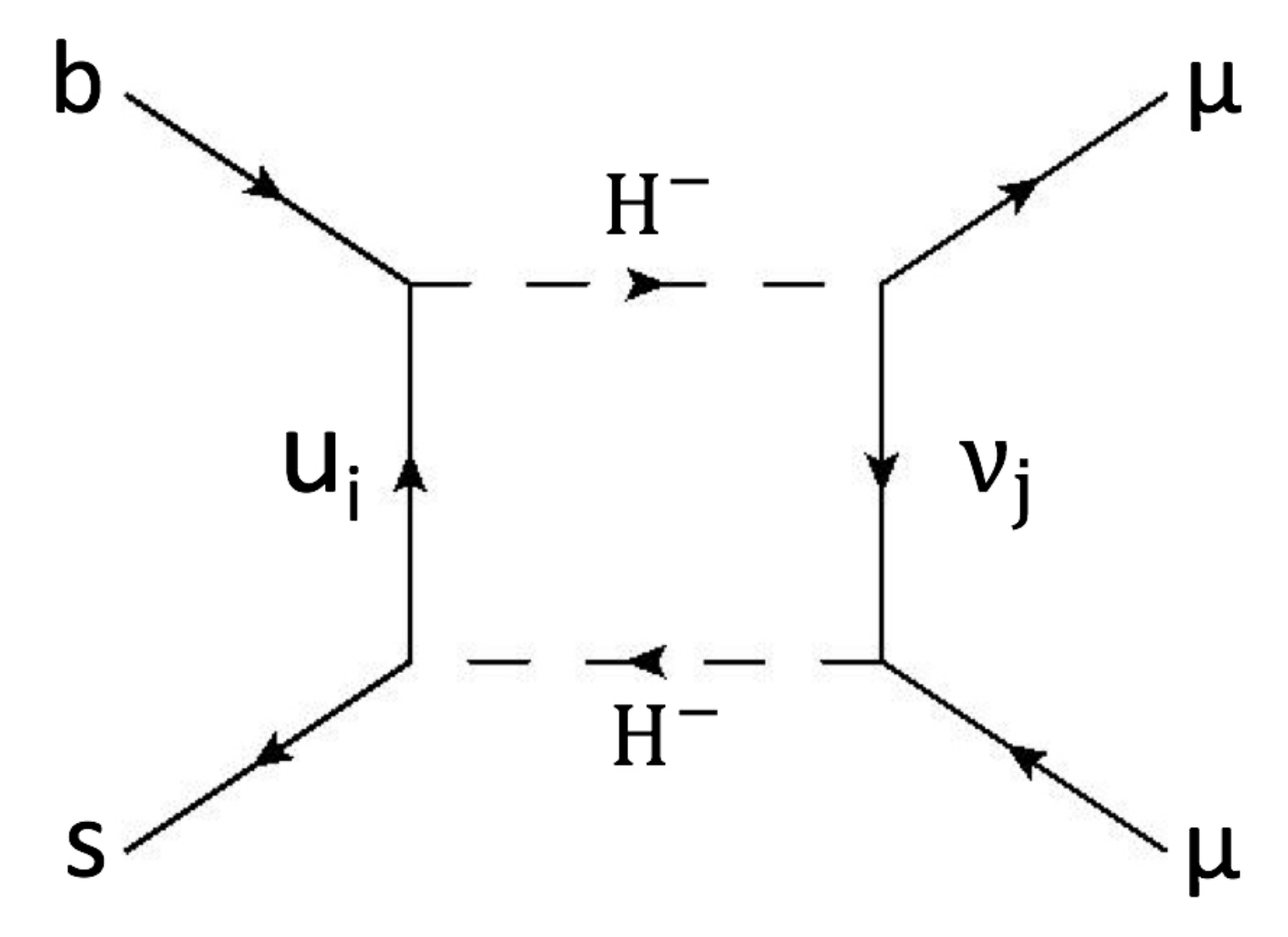}
\end{center}
\caption{ Diagram that contributes to the $B \to K \mu\mu$ in case (B) and case $(C)$.}
\label{BtoKmumu-box}
\end{figure}

In Fig. \ref{scenario3}, we investigate 
the sizes of $\Delta C^\mu_9$ and $ \Delta C^\mu_{10}$, setting $\rho_e^{\tau \mu}=1$, $0.1$ and $(m_A, \, m_H, \, m_{H^\pm})=(200 \, {\rm GeV}, \, 250 \, {\rm GeV}, \, 200 \, {\rm GeV})$. $\rho_e^{\mu \tau}$ is fixed at  $\rho_e^{\mu \tau}=-0.034, \, -0.34$ that correspond to $\delta\alpha_\mu=2.61 \times10^{-9}$.
In the plots, the $\rho_u^{tt}$ and $\rho_u^{ct}$ dependences are shown, to see the contribution of the box diagram in Fig. \ref{BtoKmumu-box}. $\rho_u^{tc}$ is vanishing on the both panels.
The gray region is excluded by the $B_s-\overline{B_s}$ mixing (red lines) and  $\tau \to  \mu\gamma$ process (dotted green lines). The dashed green lines and dashed purple lines denote the predictions of $\Delta C^\mu_9$ and $\Delta C^\mu_{10}$ for the each case.

In this case, the deviations of $\Delta C^\mu_9$ and $\Delta C^\mu_{10}$ can be sizable, according to the diagrams in Fig. \ref{BtoKmumu-penguin} and Fig. \ref{BtoKmumu-box}. In particular, the box diagram in Fig. \ref{BtoKmumu-box} can lead
the flavor universality violation in the $B \to K^{(*)} ll$ processes.
In the case (B), however, the box diagram in Fig. \ref{BtoKmumu-box} predicts two muons in the final state to be right-handed, so that the relation, $\Delta C^\mu_9=\Delta C^\mu_{10}$, is predicted. According to the recent global analyses \cite{Altmannshofer:2017yso,DAmico:2017mtc}, $\Delta C^\mu_9=-\Delta C^\mu_{10}$ is favored. 
$R(K)$ is, in fact, estimated as $R(K)=1 +0.23 \Delta C^\mu_9- 0.233 \Delta C^\mu_{10}$ in $1$ GeV$^2$ $\leq q^2 \leq 6$ GeV$^2$ \cite{Celis:2017doq},
so that the relation, $\Delta C^\mu_9=\Delta C^\mu_{10}$, leads $R(K)$ to almost unit.
Thus, we conclude that it is difficult to achieve the explanations of the $R(K^{(*)})$ anomaly in the case (B). 
Such a positive $\Delta C_{10}$ is disfavored by $B_s \to \mu \mu$.
As mentioned above, it is also difficult that the explanation of $R(D)$ is compatible with the one of $(g-2)_{\mu}$,
because of the constraint from the lepton universality of $B \to D^{(*)} l \nu.$
Note that $\Delta C^\mu_9$ is small on this plane in Fig. \ref{scenario3}. If $\rho_u^{tc}$ is not vanishing,
sizable $\Delta C^\mu_9$ can be derived as shown in Fig. \ref{scenario1}, although the $\Delta C^\mu_9$
is flavor universal. Then, it is possible that we explain both the $R(D)$ and $P^\prime_5$ anomalies by the one parameter set, but $R(K^{(*)})$ is not compatible with the explanation.

\subsubsection{Case (C): $\rho_e^{ij} = 0$ and $(\widetilde{\rho_\nu})^{j \mu } \neq 0$}
\label{sec;caseC}

\begin{figure}[t]
  \begin{center}
 \includegraphics[width=0.4\textwidth]{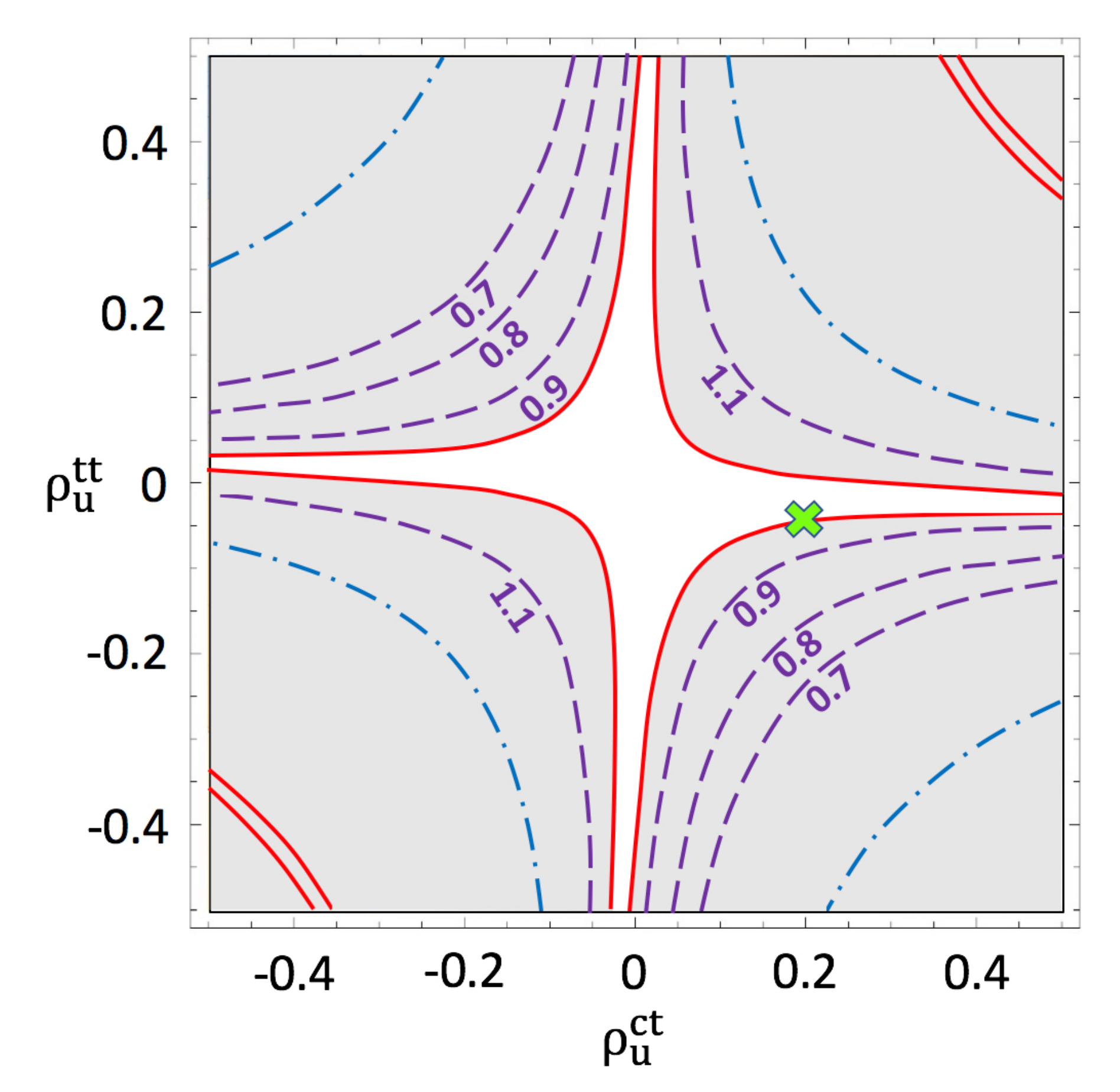}
 \includegraphics[width=0.4\textwidth]{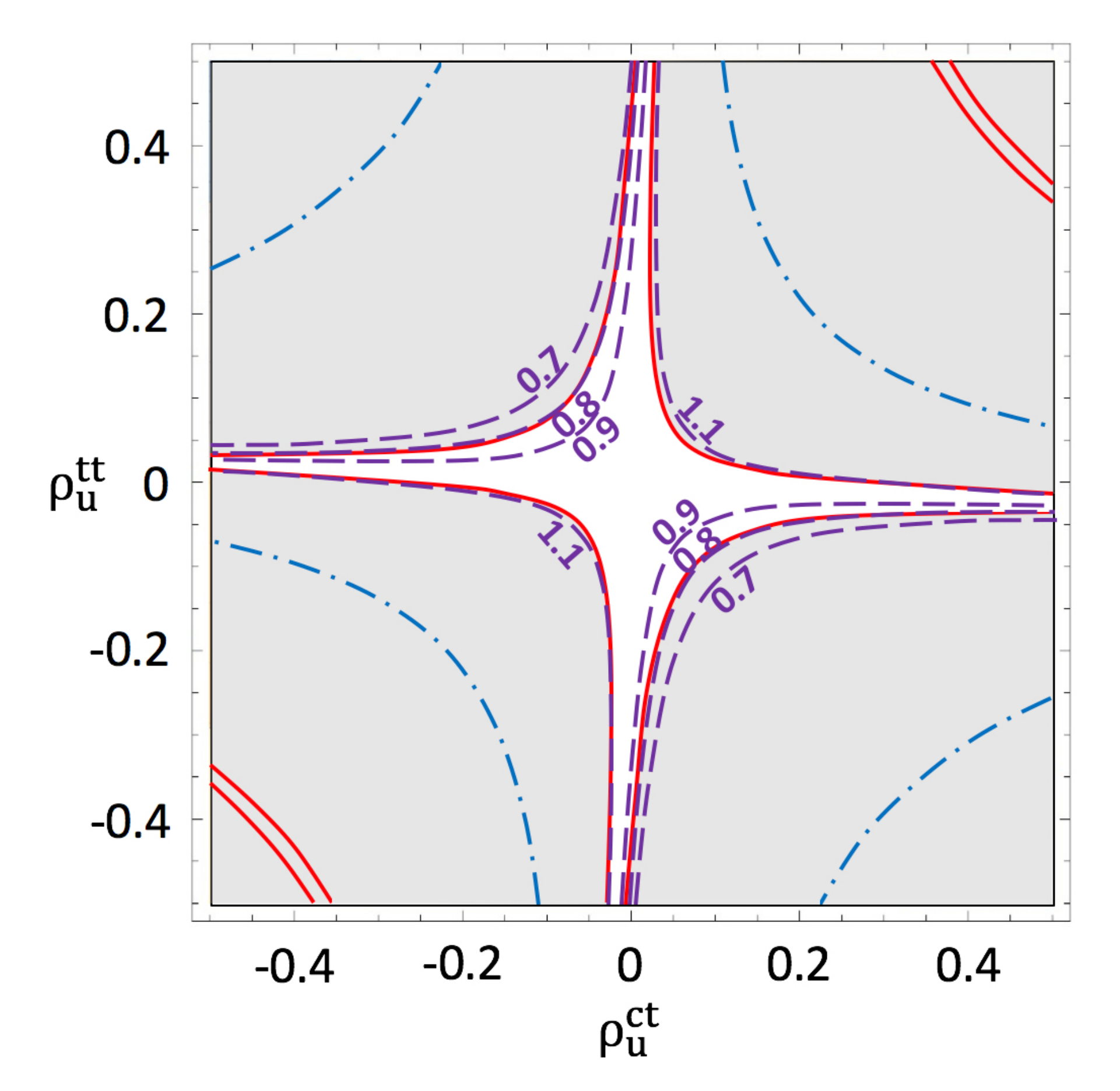}
        \caption{$\rho_u^{tt}$ vs. $\rho_u^{ct}$  in the case (C) with $\overline{\rho_\nu}=1(\rm{left}), \, 2(\rm{right})$ and $(m_A, \, m_H, \, m_{H^\pm})=(200 \, {\rm GeV}, \, 200 \, {\rm GeV}, \, 200 \, {\rm GeV})$. 
         The gray region is excluded by the $B_s-\overline{B_s}$ mixing (solid red lines) and $b \to s \gamma$ (dotted-dashed blue lines). The dashed purple lines denote the predictions of $R(K)$.}
    \label{scenario4}
  \end{center}
\end{figure}
Finally, we study the case (C). 
The all elements of $\rho_e$ are vanishing and some elements of $\rho_\nu$ are sizable in this case. 
As discussed in Sec. \ref{sec;rhoe}, the LFV processes strictly constrain $(\widetilde{\rho_\nu})^{ij}$, and
then we assume that the only sizable element is $(\widetilde{\rho_\nu})^{\mu j}$.
This assumption principally forbids the flavor violating processes.
$(\widetilde{\rho_\nu})^{\mu j}$ is also constrained by the (semi)leptonic $B$ decays, as shown in Sec. \ref{Btolnu} and Sec. \ref{BtoDlnu}, when $\rho_u^{tc}$ is large.
Let us define the following parameter,
\beq
\overline{\rho_\nu}= \sqrt{ \sum_j \left |(\widetilde{\rho_\nu})^{\mu j } \right |^2},
\eeq
and draw Fig. \ref{scenario4} fixing $\overline{\rho_\nu}=1, \, 2$ on the left and right panels, respectively. 

Based on Ref. \cite{Celis:2017doq}, we evaluate $R(K)$, that is the ratio 
between BR($B^+ \to K^+ \, \mu \mu$) and BR($B^+ \to K^+ \,ee$). 
$R(K)$ is reported 
in each bin of $q^2$ GeV$^2$, which is the invariant mass of two leptons in the final state \cite{Aaij:2014ora}. 
In particular, the result in $B^+ \to K^+ \, \mu \mu$ with $1$ GeV$^2$ $\leq q^2 \leq 6$ GeV$^2$ is smaller than the SM predictions: $R(K)=0.745^{+0.090}_{-0.074}\pm0.036$ \cite{Aaij:2014ora}.
The lepton universality is measured in $B_0 \to K^* \, \mu \mu$ as well, and the experimental result also shows the similar sign about the lepton universality violation \cite{LHCbnew}.

In our model, $R(K)$ is deviated by the diagram in Fig. \ref{BtoKmumu-box} via the leptonic Yukawa couplings. 
In the case (C), the leptons in the final state can be left-handed, so that $\Delta C_9=-\Delta C_{10}$ is predicted.
In Fig.  \ref{scenario4}, the predicted $R(K)$ is drawn by the dashed purple lines. The number on the each line
corresponds to the size of $R(K)$. The relevant parameters are fixed at $\overline{\rho_\nu}=1(\rm{left~panel}), \, 2(\rm{right~panel})$ and $(m_A, \, m_H, \, m_{H^\pm})=(200 \, {\rm GeV}, \, 200 \, {\rm GeV}, \, 200 \, {\rm GeV})$.
 The gray region is excluded by the $B_s-\overline{B_s}$ mixing (solid red lines) and $b \to s \gamma$ (dotted-dashed blue lines).
As we see in Fig. \ref{scenario4}, large $\overline{\rho_\nu}$ is required even in the light charged Higgs scenario. The strongest constraint comes from $B_s-\overline{B_s}$ mixing, and then
$R(K)$ can reach $0.8$, that is within $1 \sigma$ region, when $\overline{\rho_\nu}=2$ and $m_{H^{\pm}}=200$ GeV.

In such a case with large $\overline{\rho_\nu}$, the cosmological observations and the neutrino experiments will severely 
constrain our model. Let us simply assume that the active neutrinos consist of right-handed and left-handed neutrinos:
they are Dirac neutrinos. In the case (C), the coupling with muon, $\widetilde{\rho_\nu}^{\mu i}$, is large and the others are small. This means that the only one right-handed neutrino that couples to muon is introduced effectively.
In our scenario, the right-handed neutrino interacts with the SM particles through the $\overline{\rho_\nu}$ coupling,
and it is in the thermal equilibrium up to a few MeV, when $\widetilde{\rho_\nu}^{\mu i}$ is ${\cal O}(1)$.
The effective number, $N_{eff}$, of neutrinos in our universe is measured by the Planck experiment: $N_{eff}=3.36 \pm 0.34$ (CMB only) \cite{Ade:2013zuv}.
If the decoupling temperature of the right-handed neutrino is small, $N_{eff}$ could be estimated as
$N_{eff} \approx 4$, that is excluded by the recent cosmological observation. 
In order to raise the decoupling temperature and decrease $N_{eff}$, $\widetilde{\rho_\nu}^{\mu i}$ may be required to be
less than ${\cal O}(0.1)$ \cite{Davidson:2009ha}. 

The right-handed neutrino, on the other hand, is not needed to be an active neutrino,
in our setup. In Fig. \ref{scenario4}, the right-handed neutrino mass is vanishing, but the result would not be modified so much even if the small Majorana mass of the right-handed neutrino is introduced.
Let us define the right-handed neutrino that couples to muon as $\nu^{1}_R$. Then, the relevant terms are given by
\begin{equation}
\bar{L}_L^i (V_{\nu})^{ij} \widetilde H_1 y^j_\nu \nu_R^i +m_R \overline{\nu_R^{1 \, c} }\nu_R^1+ \widetilde \rho^{ \mu 1}_\nu \overline{L^\mu_L} \widetilde H_2 \nu^{1}_R+h.c..
\end{equation}
Here, $ y^1_\nu$ can be assumed to be vanishing without conflict with the neutrino observables.
As far as $H_2$ does not develop non-vanishing VEV, $ \widetilde \rho^{ \mu 1}_\nu$ does not contribute to the masses of the active neutrinos, even if $m_R$ is sizable. The decay of $\nu^{1}_R$ may be suppressed according to 
the alignment of $ \rho_\nu$. It would be interesting to discuss the compatibility between the dark matter abundance and
$R_K$, as discussed in Ref. \cite{Kawamura:2017ecz}. In our case, $\nu^{1}_R$ can decay to leptons through $\rho^{ij}_\nu$\footnote{$\widetilde \rho^{ i 2}_\nu$ and $\widetilde \rho^{ i 3}_\nu$ are negligibly small, but not vanishing.}, as far as 
$\nu^{1}_R$ is heavier than $\nu^{2}_R$ and $\nu^{3}_R$, that decouple with the thermal bath above the QCD phase transition temperature.\footnote{Recently, the model with light $\nu_R$ that strongly couples to leptons is discussed, motivated by the $R(D^{(*)})$ anomaly \cite{Asadi:2018wea,Greljo:2018ogz}.}

The neutrino scattering with nuclei also strongly constrains our model.
The relevant process is the neutrino trident production: $\nu N \to \nu \mu \overline{\mu}$ \cite{Brown:1973ih}.
In our model with sizable $\widetilde \rho_\nu^{\mu 1}$, the charged Higgs exchanging 
enlarges the cross section but the contribution does not interfere with the SM correction,
so that the prediction is not deviated from the SM prediction so much.
$\widetilde \rho_\nu^{\mu 1}$, however, is very large to violate the lepton universality of $B \to K^{(*)} ll$,
so that we obtain the limit on the deviation of $R_K$ and $R_{K^{*}}$.
When $m_{H^{\pm}}$ is set to 200 GeV, the upper bound on $\widetilde \rho_\nu^{\mu 1}$ is about $1$ to avoid 
the $2 \sigma$ deviation of the experimental result \cite{Altmannshofer:2014pba}.
Thus, the $\widetilde \rho_\nu^{\mu 1} \approx 2$ scenario is totally excluded, as far as $m_R$ is not introduced.
 
We conclude that the scenario with large $\nu^{1}_R$ coupling is excluded by the cosmological observations and the neutrino experiments, if $\nu^{1}_R$ is a part of the active neutrinos. We can easily introduce the mass term of 
$\nu^{1}_R$, i.e. $m_R$, since $\nu^{1}_R$ is neutral under the SM gauge symmetry.
Then, the bound from the trident production can be evaded, since $\nu^{1}_R$ is not an active neutrino in this case.
When small other elements of $( \widetilde \rho_\nu)^{i 2}$ and $( \widetilde \rho_\nu)^{i 3}$ are allowed and $\nu^{1}_R$ is heavier than $\nu^{2,3}_R$, $\nu^{1}_R$ can decay to the SM leptons in association with $\nu^{2,3}_R$.
$\nu^{2,3}_R$ can be interpreted as the active neutrinos, if the Majorana masses of $\nu^{2,3}_R$ are vanishing.
Then, $( \widetilde \rho_\nu)^{i j}$, except for $( \widetilde \rho_\nu)^{\mu 1}$, should be smaller than ${\cal O}(0.1)$.

If the decay of $\nu^{1}_R$ is much suppressed, the abundance of $\nu^{1}_R$ would be constrained by
the cosmology. The cold dark matter case is similar to the result in Ref. \cite{Kawamura:2017ecz}. 
In this paper, the consistency with the cosmological observation in such a dark matter case is beyond our scope.
In Sec. \ref{sec;collider}, we propose the direct search for $\nu^{1}_R$ at the LHC.


\subsection{Summary of the capabilities to explain the excesses}

We summarize the possibility that our model can explain the 
excesses in the flavor physics, choosing the proper parameter set.
In Table \ref{table9}, our conclusion about the each excess is shown.
On the first, second and third rows, $\rho_u^{tt}$, $\rho_u^{tc}$ and $\rho_u^{ct}$ are only sizable in the case (B) and (C), respectively.
The each column corresponds to the capability to explain the each excess denoted on the top row.
The symbol, ``$\bigcirc$", means that
our predictions are within the 1$\sigma$ regions of the experimental results.
In the box with  ``$\times$", our prediction is out of the 2$\sigma$ region.
In the box with ``$\triangle$", 
the predictions can be within the 2$\sigma$ region of the experimental results, i.e., $P^\prime_5$ and $R(K)=0.745^{+0.090}_{-0.074}\pm0.036\simeq0.745^{+0.097}_{-0.082}$($q^2$ [1, 6]GeV$^2$) \cite{Aaij:2014ora},
if $\overline{\rho_\nu}$ is ${\cal O}(1)$. The Dirac neutrino case predicts $N_{eff} \approx 4$ and
is in tension with the recent cosmological observation. The neutrino trident production also excludes the case with 
$\overline{\rho_\nu} >1$. We can also introduce the small Majorana mass term, $m_R$, to decrease $N_{eff}$.

\begin{table}[h]
  \begin{center}
    \begin{tabular}{|c|c|c|c|c|c|}
      \hline
        &$R(K^{(*)})$ & ~~$P_5'$~~ & ~$R(D)$~&$R(D^*)$&~~$\delta\alpha_\mu$~~\\
\hline \hline
 \multicolumn{6}{|c|}{(B) $\rho_e \neq 0$, $\rho_\nu =0$ }\\
\hline \hline
$\rho_u^{tt}$&$\times$&$\times$&$\times$&$\times$&$\bigcirc$ \\
\hline
$\rho_u^{tc}$&$\times$&$\bigcirc$&$\bigcirc$&$\times$&$\times$ \\
\hline
$\rho_u^{ct}$&$\times$&$\times$&$\times$&$\times$&$\bigcirc$ \\
\hline \hline
\multicolumn{6}{|c|}{(C) $\rho_e =0$, $\rho_\nu \neq0$}\\
\hline \hline
$\rho_u^{tt}$&$\triangle$&$\triangle$&$\times$&$\times$&$\times$ \\
\hline
$\rho_u^{tc}$&$\times$&$\bigcirc$&$\bigcirc$&$\times$&$\times$ \\
\hline
$\rho_u^{ct}$&$\triangle$&$\triangle$&$\times$&$\times$&$\times$ \\
\hline
    \end{tabular}
    \caption{Summary of the capabilities to explain the excesses. In the each observable, 
  our prediction is evaluated by the symbols, ``$\bigcirc$", ``$\triangle$" and ``$\times$".
  The meanings are explained in the text. }
    \label{table9}
  \end{center}
\end{table}

In the end, it is difficult to explain all of the excesses in our parameterization. 
The explanations of $P^\prime_5$ and $R(D)$ can be done by the sizable $\rho^{tc}_{u}$ and the $\rho_e^{\mu \tau}$,
but cannot be compatible with the solutions to the $(g-2)_{\mu}$ and $R(K^{(*)})$ anomalies.
This is because the charged Higgs that couple to $b$, $c$ and $\mu$ largely violate the
lepton universality of $B \to D^{(*)} l \nu$.

\section{Our signals at the LHC}
\label{sec;collider}

Before closing our paper, we discuss the possibility that our 2HDM is tested by the LHC experiments.
In our scenarios, the extra scalars are relatively light: we fix the masses at 200 GeV or 250 GeV.
Thus, the main targets to prove our model are the direct signals originated from the scalars.

In the case (A), there are Yukawa couplings between the scalars and heavy quarks, denoted by $\rho_u^{tc}$, $\rho_u^{ct}$ and $\rho_u^{tt}$. If either $\rho_u^{tc}$ or $\rho_u^{ct}$ is ${\cal O}(1)$, we obtain large $\Delta C_9$, that can explain the $P^\prime_5$ excess. In this case, the neutral and charged scalars are produced in
association with top quark or bottom quark in the final state.
The produced scalars dominantly decay to heavy quarks, so that there are tt/bb/tb quarks in the final state. 
Such a case has been studied in Ref. \cite{Iguro:2017ysu}. \footnote{See also Refs. \cite{Kim:2015zla,Kohda:2017fkn,Gori:2017tvg,Chiang:2015cba,Gori:2016zto,Atwood:2013xg,Craig:2015jba,Goldouzian:2014nha,Patrick:2016rtw,Patrick:2017ele,Arhrib:2017veb,Altmannshofer:2017poe,Campos:2017dgc}.}

\begin{table}[t]
  \begin{center}
    \begin{tabular}{|c|c|c|}
      \hline
      $\sqrt s$&13TeV&8TeV\\
\hline \hline
\multicolumn{3}{|c|}{$m_{H^\pm}=$ 200 [GeV] }\\
\hline
$\sigma(b+\overline{c}\to H^\pm)$&792$\times |\rho_u^{tc}|^2$&287$\times|\rho_u^{tc}|^2$\\
\hline
$\sigma(g+s\to t+H^-)$&11.4$\times|\rho_u^{ct}|^2$&3.0$\times|\rho_u^{ct}|^2$\\
\hline
$\sigma(g+g\to \overline{s}+t+H^-)$&4.0$\times|\rho_u^{ct}|^2$&0.88$\times|\rho_u^{ct}|^2$\\
\hline \hline
\multicolumn{3}{|c|}{$m_{\phi}=$ 200 [GeV] ( $\phi= H, \, A$ )  }\\
\hline
$\sigma(g+c\to t+\phi)$&3.8$\times|\rho_u^{ct}|^2$&0.92$\times|\rho_u^{ct}|^2$\\
\hline
$\sigma(g+g\to \overline{c}+t+\phi)$&1.36$\times|\rho_u^{ct}|^2$&0.3$\times|\rho_u^{ct}|^2$\\
\hline \hline
\multicolumn{3}{|c|}{$m_{\phi}=$ 250 [GeV] }\\
\hline
$\sigma(g+c\to t+\phi)$&0.84$\times|\rho_u^{ct}|^2$&0.17$\times|\rho_u^{ct}|^2$\\
\hline 
$\sigma(g+g\to \overline{c}+t+\phi)$&2.4$\times|\rho_u^{ct}|^2$&0.55$\times|\rho_u^{ct}|^2$\\
\hline
    \end{tabular}
    \caption{Heavy Higgs Production cross section in pb. We added normal and conjugate cross sections, just as adding $\sigma$($g+s\to t+H^-$) and $\sigma$($g+\overline{s}\to \overline{t}+H^+$) and denote as $\sigma$($g+s\to t+H^-$).}
    \label{HHP}
  \end{center}
\end{table}

\begin{figure}[H]
\begin{center}
\includegraphics[width=4.5cm]{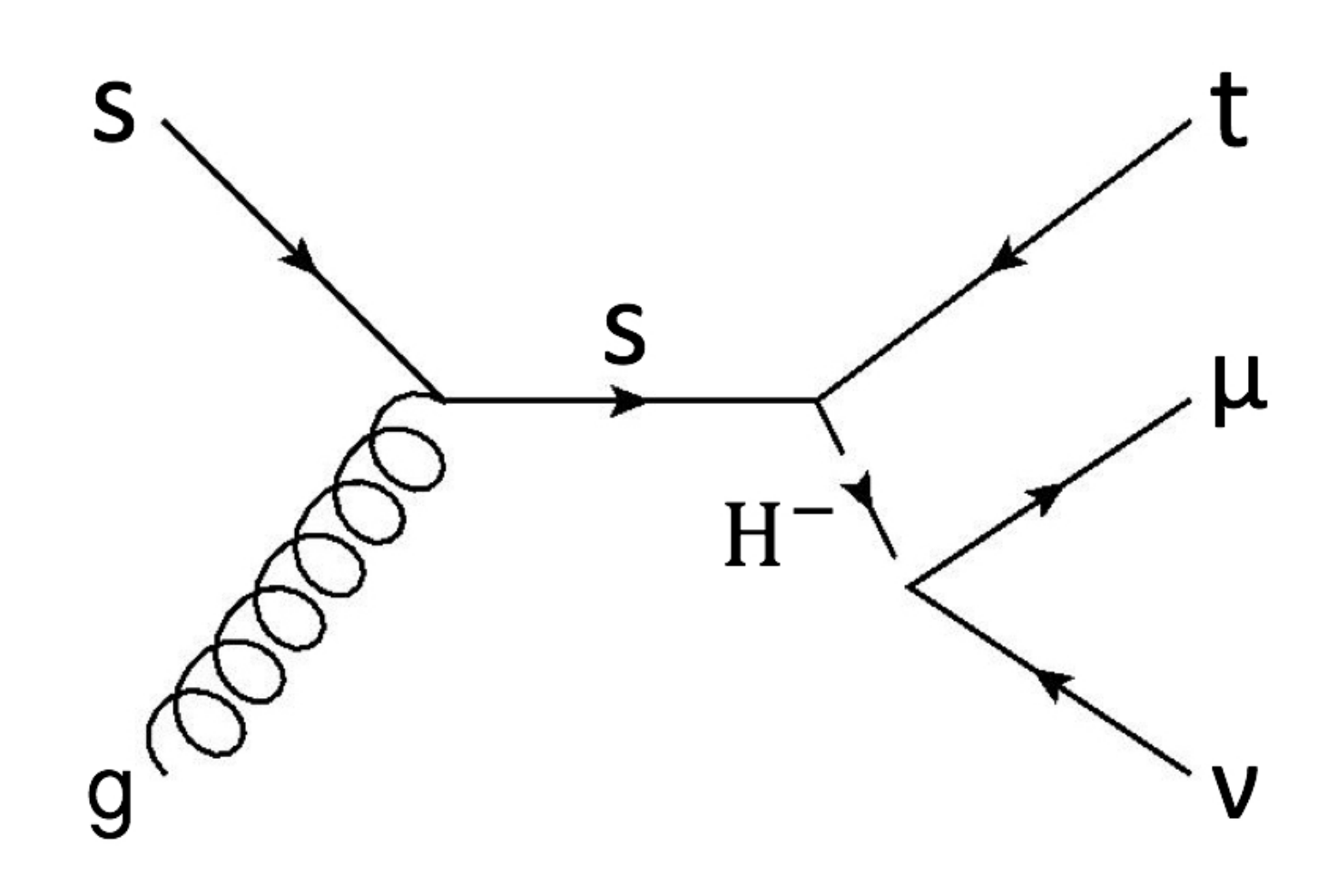}
\includegraphics[width=4.5cm]{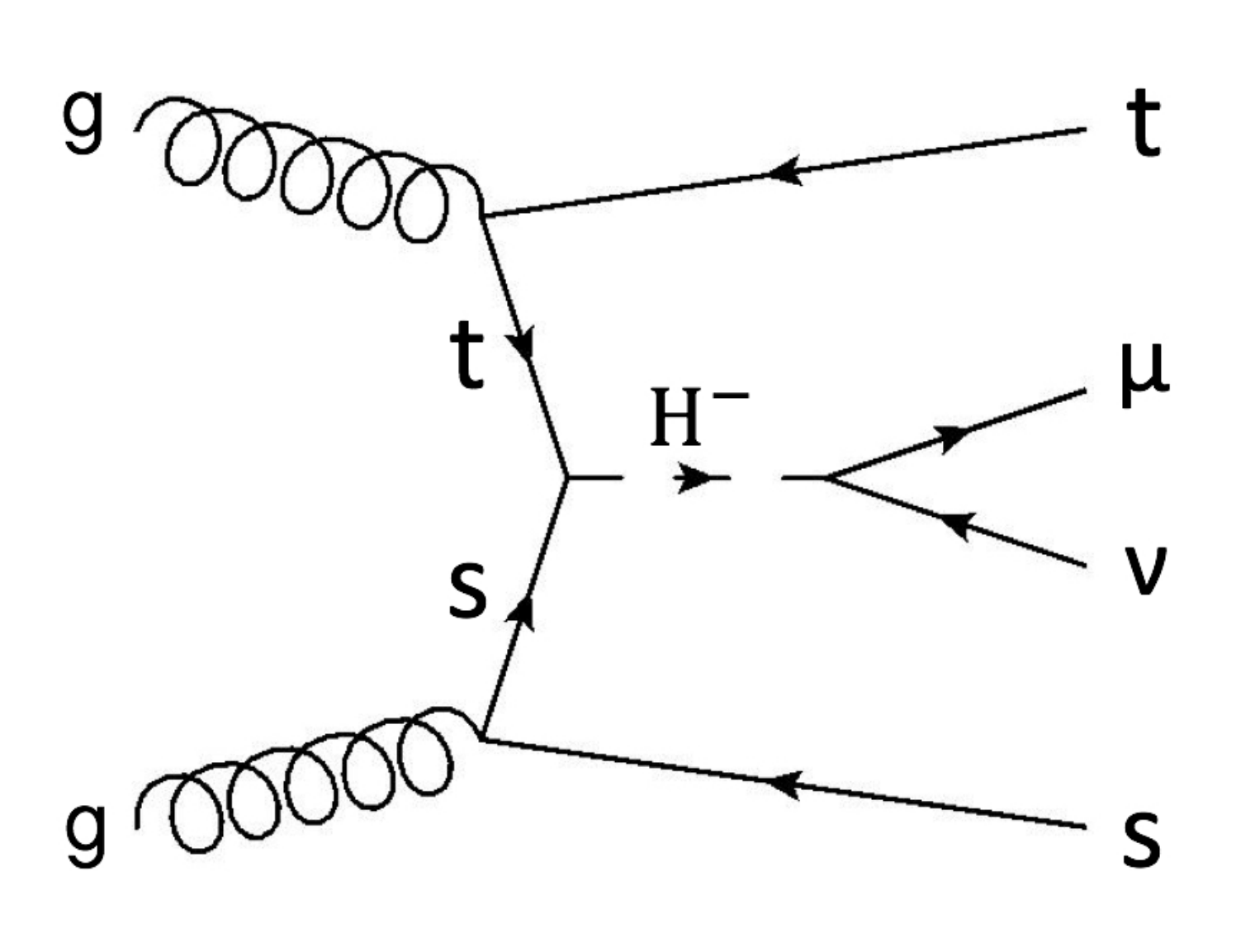}
\end{center}
\caption{ Diagrams that contributes to the $\mu\nu$ resonance.}
\label{munu}
\end{figure}

In the case (B),  the neutral scalars can decay to $\mu$ and $\tau$, and
the charged Higgs decays to $\mu$ or $\tau$ with one neutrino.
The scalars are produced via $\rho_u^{ct}$ coupling, and then 
the production cross sections of the scalars at the LHC with $\sqrt s=13$ TeV (8 TeV) are estimated 
in Table \ref{HHP}, using CALCHEP\cite{Belyaev2012qa}. Note that cteq6l1 is applied to the parton distribution function.
Here, we quantitatively study our signal on the benchmark points in Fig. \ref{scenario3}. We put the green x-marks on the figures.
On the benchmark point (B1), the parameters are aligned as
\begin{eqnarray}
m_{H^\pm}&=&m_A=200 \, {\rm GeV},~m_H=250 \, {\rm GeV},  \nonumber \\
(\rho_u^{tt}, \, \rho_u^{ct})&=&(0.005, \, 0.2),  \nonumber \\
(\rho_e^{\tau \mu}, \, \rho_e^{\mu \tau})&=&(1, \, -0.0341).  
\end{eqnarray}
This parameter set leads a sizable deviation of $(g-2)_\mu$: $\delta\alpha_\mu=2.61\times10^{-9}$.
On this point, the charged Higgs mainly decays to $\mu \nu$ through the diagram in Fig. \ref{munu} and the heavy neutral scalar decays to $\mu \tau$: 
\begin{eqnarray}
BR(H^- \to \mu \bar{\nu}) &\approx& 99.3 \%,~BR(H^- \to \overline{t} s) \approx 1 \%, \nonumber \\
BR(A \to \mu \tau) &\approx& 99.3 \%,~BR(A \to t c) \approx 0.7 \%, \nonumber \\
BR(H \to \mu \tau) &\approx& 96.9 \%,~BR(H \to t c) \approx 3.1 \%.
\end{eqnarray}
Following Table \ref{HHP}, the production cross section of the charged Higgs is
estimated as 2.46 pb at the LHC with $\sqrt s=13$ TeV. The search for a new heavy resonance
decaying to $e/\mu$ and neutrino has been developed recently \cite{Aaboud:2017efa} and the upper bound on the production is about 0.6 pb, that naively leads the upper bound on $|\rho_u^{ct}|$ as $ |\rho_u^{ct}| \lesssim 0.2$. In our model, however, there are top quarks in the final state, so that
the top quark will make the signals fuzzy.
The search for a new resonance decaying to $\tau \, \nu$/ $\tau \, \mu$ is also attractive, because the decay is
predicted by the charged Higgs and the neutral Higgs. 
It is challenging and actually the heavy mass region is surveyed by the ATLAS \cite{Aaboud:2018vgh} and CMS collaborations \cite{Khachatryan:2015pua,CMS-Wprime}.
As discussed in Sec. \ref{BtoDlnu}, the excesses in $B \to D^{(*)} \tau \nu$ require rather large Yukawa couplings,
so that we expect that the direct search for the resonance at the LHC can reach the favored parameter region near future.
The detail analysis is work in progress.

On the benchmark point (B2), the parameters are fixed at
\begin{eqnarray}
m_{H^\pm}&=&m_A=200 \, {\rm GeV},~m_H=250 \, {\rm GeV},  \nonumber \\
(\rho_u^{tt}, \, \rho_u^{ct})&=&(0.006, \, 0.2),  \nonumber \\
(\rho_e^{\tau \mu}, \, \rho_e^{\mu \tau})&=&(0.1, \, -0.341).  
\end{eqnarray}
Then, the sizable deviation of $(g-2)_\mu$ is estimated as $\delta\alpha_\mu=2.61\times10^{-9}$.
Since $\rho_e^{\mu \tau}$ is sizable, the charged Higgs decays to $\tau \nu$:
\begin{eqnarray}
BR(H^- \to \mu \bar{\nu}) &\approx& 7.5 \%,~BR(H^- \to \tau \bar{\nu}) \approx 86.9 \%,~BR(H^- \to \overline{t} s) \approx 5.6 \%, \nonumber \\
BR(A \to \mu \tau) &\approx& 94.4 \%,~BR(A \to t c) \approx 5.6 \%, \nonumber \\
BR(H \to \mu \tau) &\approx& 79.6 \%,~BR(H \to t c) \approx 20.4 \%.
\end{eqnarray}
In this case, the charged Higgs mainly decays to $\tau \nu$, and can evade the 
bound from the $\mu \nu$ resonance search.

\begin{figure}[h]
\begin{center}
\includegraphics[width=4.5cm]{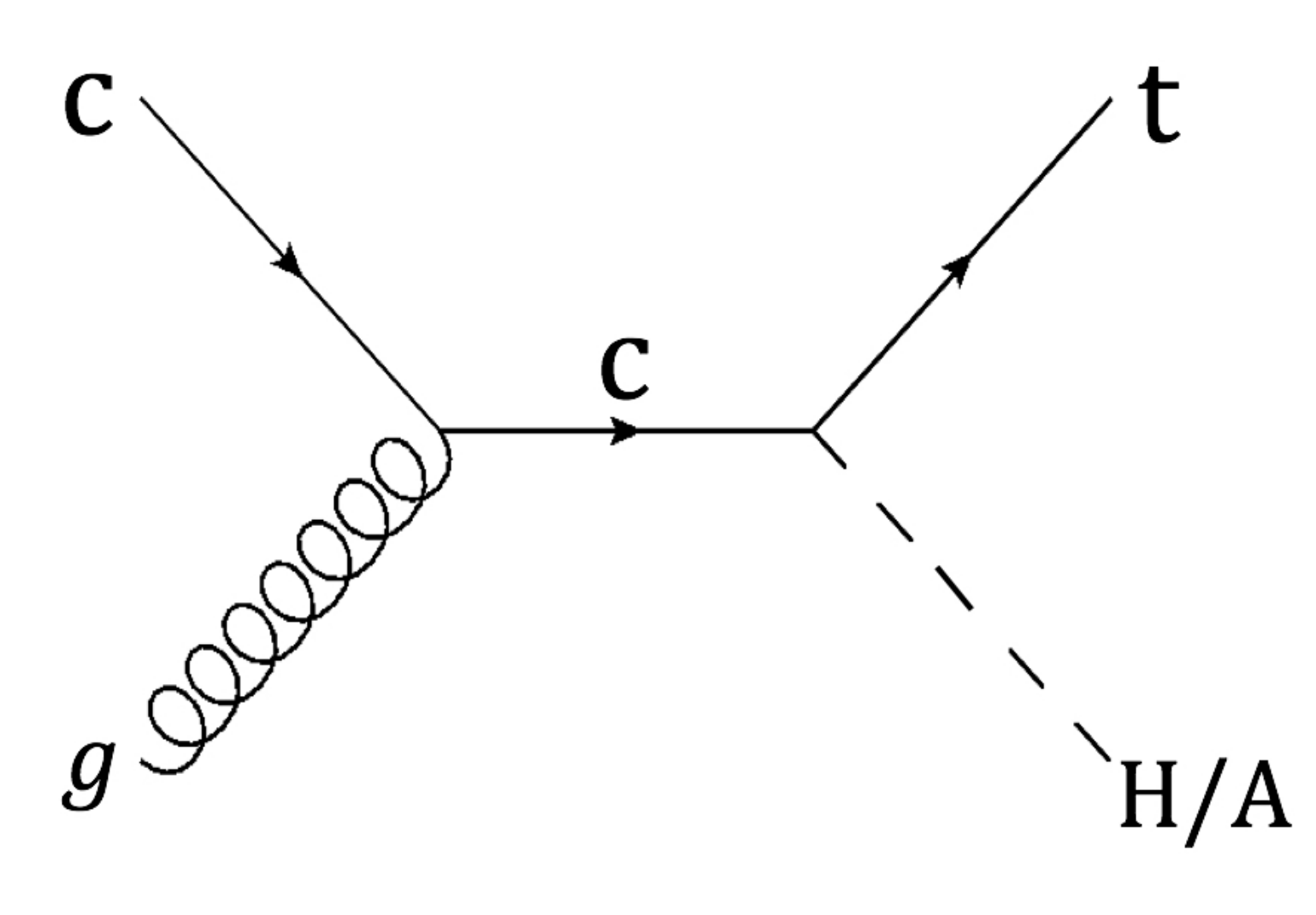}
\end{center}
\caption{Diagram that contributes to our monotop.}
\label{monotop}
\end{figure}

In the case (C), the scalars are produced due to the large $\rho_u^{ct}$. 
The produced neutral scalars decay to two neutrinos in this case, so that they predict the invisible signal.
The charged scalar decays to one muon and one neutrino. This signal is similar to the case (B).
On the benchmark point in Fig. \ref{scenario4}, the parameters satisfy
\begin{eqnarray}
m_{H^\pm}&=&m_A=m_H=200 \, {\rm GeV},  \nonumber \\
(\rho_u^{tt}, \, \rho_u^{ct})&=&(-0.04, \, 0.2),  \nonumber \\
\overline{\rho_\nu}^2&=&1.  
\end{eqnarray}
These parameters lead the following branching ratios,
\begin{eqnarray}
BR(H^- \to \mu \bar{\nu}) &\approx& 99 \%,~BR(H^- \to \overline{t} s) \approx 1 \%, \nonumber \\
BR(\phi_h \to \nu \nu) &\approx& 99 \%,~BR(\phi_h \to t c) \approx 1 \% ~ (\phi_h=H,  \, A).
\end{eqnarray}
The invisible decay of the heavy neutral scalars, produced by the diagram in Fig. \ref{monotop}, leads the mono-top signal:
$pp \to \phi_h t \to \nu \bar\nu t$. The current upper bound on the cross section at the LHC with $\sqrt{s}=8$ TeV is
$\sigma(pp\to t +\rm{missing})\le 0.8$ [pb] \cite{Aad:2014wza,Khachatryan:2014uma} when $m_H=200$ GeV.
Based on the results in Table \ref{HHP}, the mono-top signal on this benchmark point is about 0.3 pb, so that
it is just below the current upper bound.

\begin{figure}[h]
\begin{center}
\includegraphics[width=3.5cm]{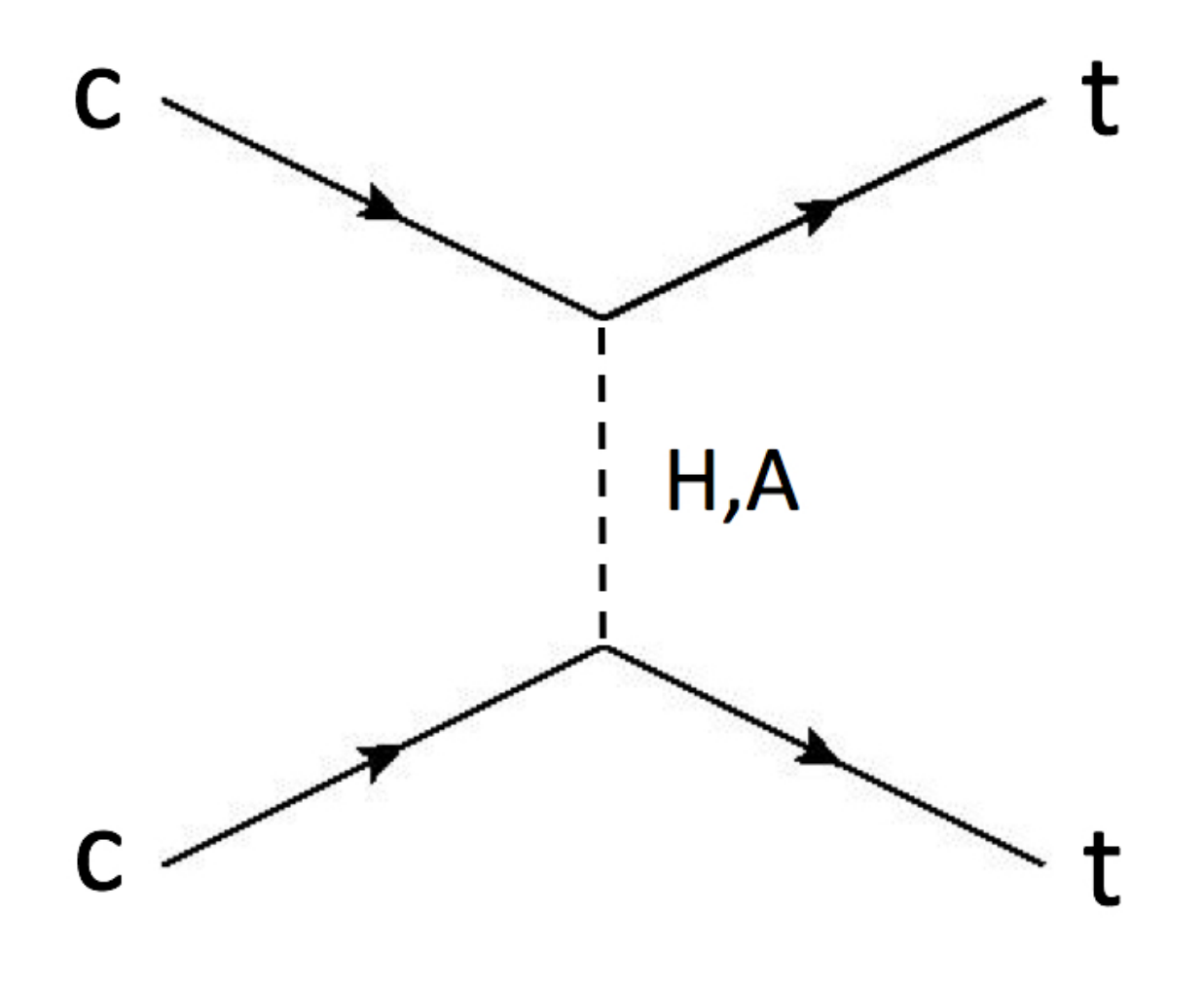}
\includegraphics[width=4.5cm]{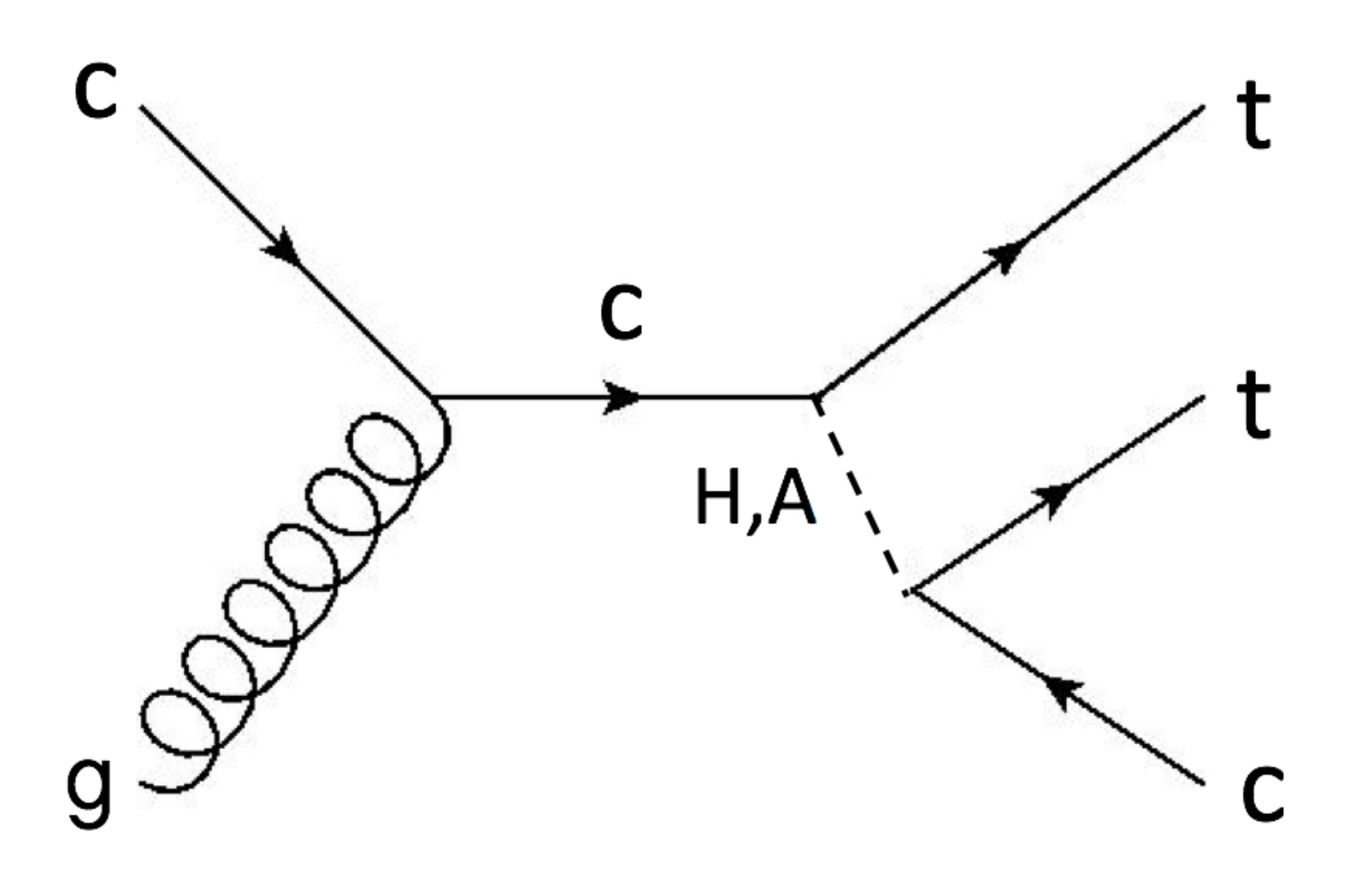}
\includegraphics[width=4.5cm]{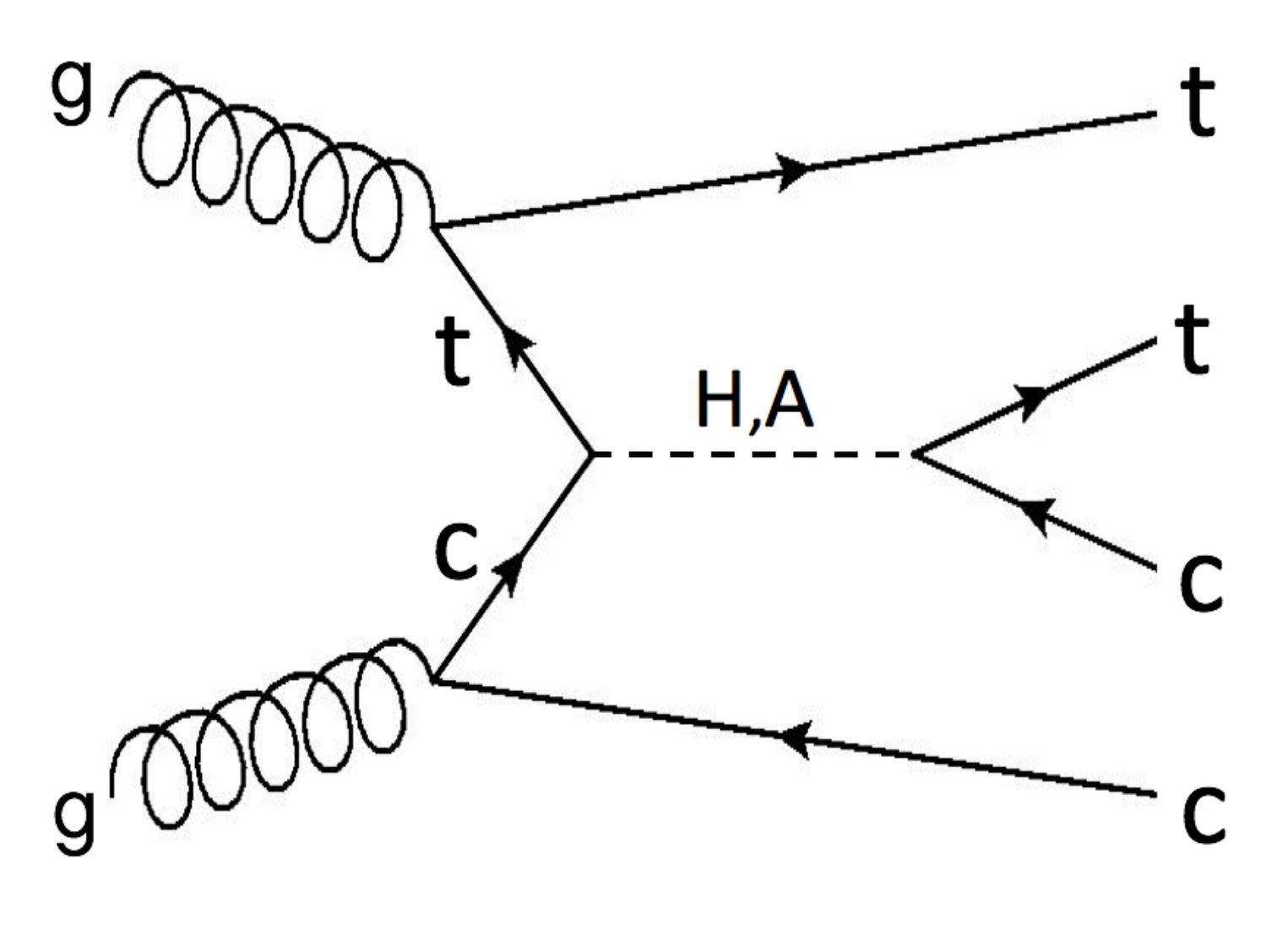}
\includegraphics[width=4.5cm]{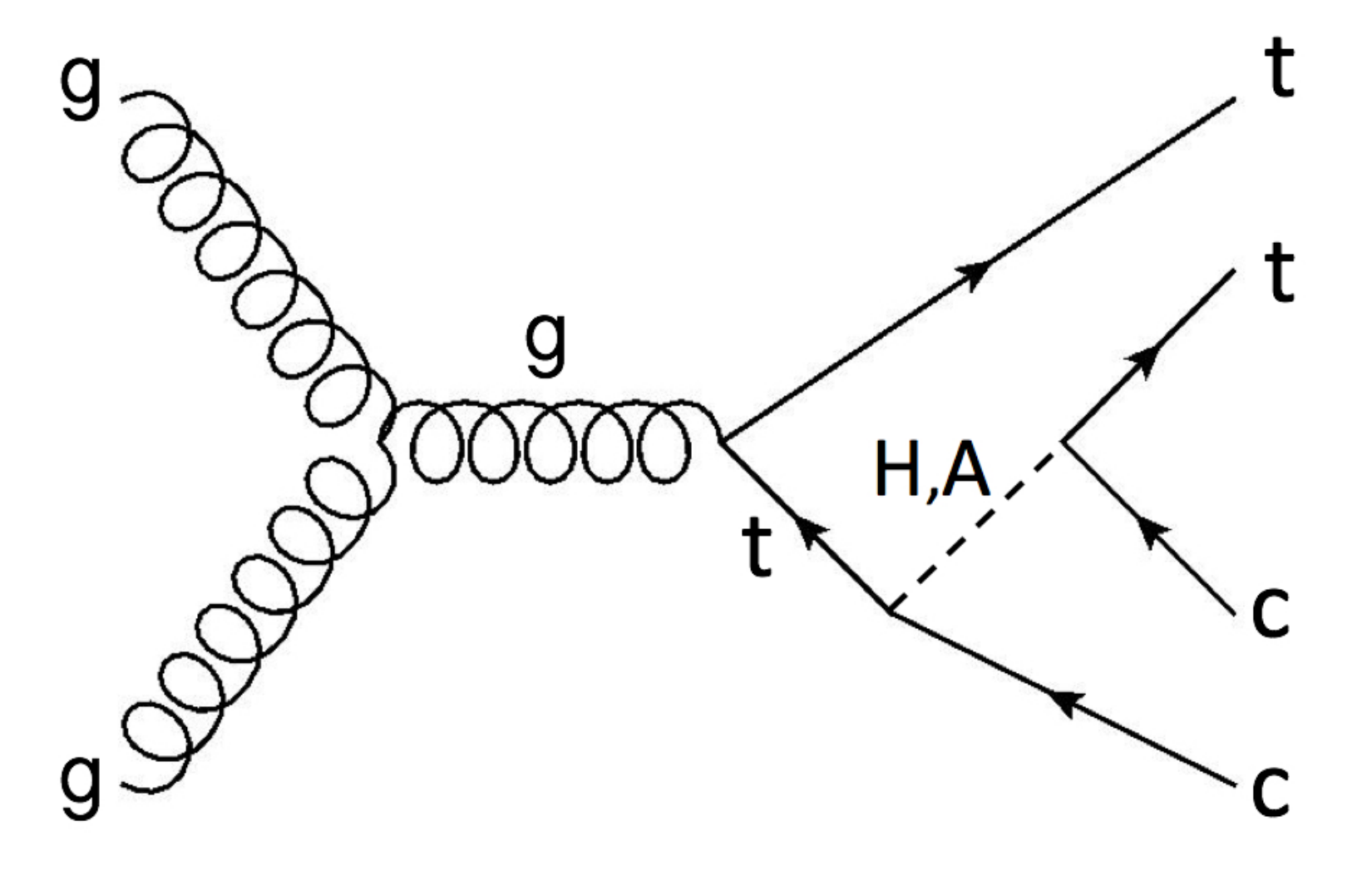}
\end{center}
\caption{Feynman diagrams relevant to the same-sign top signal. }
\label{samesigntop}
\end{figure}

In our model, the same-sign top signal is also predicted by the diagrams in Fig. \ref{samesigntop}, depending on the mass spectrum of the scalars.
If the neutral scalars, $H$ and $A$, are not degenerate, the same-sign top signal, $pp \to tt$, is enhanced by
$\rho_u^{ct}$, $\rho_u^{tc}$ couplings. The current upper bound on the cross section is 1.2 pb at the LHC with $\sqrt{s}=13$ TeV \cite{Sirunyan:2017uyt}.
When $m_A=200$ GeV and $m_H=250$ GeV, the each cross section is estimated as
\begin{align}
\sigma(pp\to tt+\bar{t}\bar{t})&=4.23\times10^{-3}|\rho_u^{tc}|^4[\rm{pb}], \nonumber\\
\sigma(pp\to tt\bar{c}+\bar{t}\bar{t}c)&=4.13\times10^{-1}|\rho_u^{tc}|^4[\rm{pb}],\nonumber\\
\sigma(pp\to tt\bar{c}\bar{c}+\bar{t}\bar{t}cc)&=1.14\times10^{-1}|\rho_u^{tc}|^4[\rm{pb}].
\label{tt}
\end{align} 
Then, our predictions on the benchmark points are below the experimental bound.
We note that the same-sign top signal is produced by the process, $pp\to tt\bar{c}+\bar{t}\bar{t}c$, rather than $pp\to cc\to tt+\bar{t}\bar{t}$, because of the production processes as shown in Fig. \ref{samesigntop}.

\section{Summary}
\label{sec;summary}
We have studied the flavor physics in type-III 2HDM.
In this model, there are many possible parameter choices, so we adopt some 
simple parameter sets motivated by the physical observables where the deviations
from the SM predictions are reported. In our scenario, the flavor violating Yukawa
couplings for up-type quarks, $\rho_u^{tc}$ and  $\rho_u^{ct}$, play an important role
in enhancing/suppressing the semileptonic $B$ decays, e.g. $B \to K^{*} ll$.
In particular, $\rho_u^{ct}$ can evade the strong bound from the flavor physics and the collider experiments,
so that $\rho_u^{ct}$ is expected to be larger than ${\cal O}(0.1)$. 
In addition, we introduce the flavor violating Yukawa couplings to the lepton sector as well: $\rho_e^{\tau \mu}$ and  $\rho_e^{\mu \tau}$. As discussed in Refs. \cite{Omura:2015nja,Omura:2015xcg},
those flavor violating couplings deviate $(g-2)_\mu$, as far as the extra
neutral scalars are not degenerate. In our paper, we have discussed the compatibility 
between the explanations of $(g-2)_\mu$, of the $B \to K^{(*)} ll$ and of the $B \to D^{(*)} \tau \nu$ excesses.
As shown in Table \ref{table9}, the explanations of $(g-2)_\mu$ and $R(D)$ require relatively large Yukawa couplings,
so the constraint from the lepton universality of $B \to D^{(*)} l \nu$ easily excludes our model. 

In order to explain the $R(K)$ excess, we need the sizable lepton flavor universality violation in the $B \to K^{(*)} ll$ processes. Then, we introduce the flavor violating Yukawa couplings involving right-handed neutrino,
and discuss the capability to explain the $R(K)$ excess in our model.
In this case, we can evade the strong experimental bounds, as far as the appropriate alignment of the Yukawa couplings
is chosen. Thus, the explanation of the $R(K)$ deviation is achieved by the box diagram involving the right-handed neutrinos
via the flavor violating neutrino Yukawa couplings. This scenario, however, can not be compatible with the other explanations, because of the stringent constraint from the lepton universality of $B \to D^{(*)} l \nu$.
In addition, the Dirac neutrino case is excluded by the recent cosmological observation. Then, the sizable
Majorana mass term for the right-handed neutrino is required to decrease the effective neutrino number.
The possible parameter choices and the capabilities of the each setup are summarized in Table \ref{table9}.

Finally, we have investigated the possibility that the LHC experiments directly test our model.
Interestingly, the direct search for new physics at the LHC can reach the parameter region that is favored by the excesses in the flavor physics  \cite{Iguro:2017ysu,Kim:2015zla,Kohda:2017fkn,Gori:2017tvg,Chiang:2015cba,Gori:2016zto,Atwood:2013xg,Craig:2015jba,Goldouzian:2014nha,Patrick:2016rtw,Patrick:2017ele,Arhrib:2017veb}.
In our scenario, the scalar are enough light to be produced by the proton-proton collider.
In the case that the charged Higgs mainly decays to one muon and one neutrino, 
the heavy resonance search at the LHC could widely cover our parameter region. 
The neutral scalar decays to two neutrinos, if the neutrino Yukawa couplings are large.
In this case, the mono-top signal could be our promising one, although the current bound has not
yet reached our parameter region. The sizable $\rho_u^{ct}$ predicts the same-sign top signal,
if the neutral scalars are not degenerate. We have confirmed that our prediction of the cross section is below the current upper bound, but we can expect that our region could be covered near future.


\section*{Acknowledgments}
The work of Y. O. is supported by Grant-in-Aid for Scientific research from the Ministry of Education, Science, Sports, and Culture (MEXT), Japan, No. 17H05404. The authors thank to Kazuhiro Tobe, Tomomi Kawaguchi, Makoto Tomoto and Yasuyuki Horii  for variable discussions. 

\appendix

\section{Various parameters for our numerical analysis}
\label{input}
Here, we summarize numerical values of various parameters we use in our numerical calculation below.
\begin{table}[H]
\begin{center}
  \begin{tabular}{|c|c|c||c|c|c|} \hline
     Quantity & Value & Refs. &  Quantity & Value & Refs.   \\ \hline \hline
     \multicolumn{3}{|c||}{CKM parameters}   & \multicolumn{3}{|c|}{parameters for hadronic matrix elements}    \\ \hline
     $\lambda$& 0.22506 &\cite{Patrignani:2016xqp}&$\rho_{D}^2$&1.128&\cite{Amhis:2016xyh}\\ 
     A&0.811~&\cite{Patrignani:2016xqp}&$\rho_{D^*}^2$&1.205&\cite{Amhis:2016xyh}\\ 
     $\bar\rho$&0.124&\cite{Patrignani:2016xqp}&$R_1(1)$&1.404&\cite{Amhis:2016xyh}\\ 
     $\bar\eta$&0.356&\cite{Patrignani:2016xqp}&$R_2(1)$&0.854&\cite{Amhis:2016xyh}\\
     \cline{1-3}
     \multicolumn{3}{|c||}{$B$ and $D$ meson parameters}   &$\Delta$&1& \cite{Tanaka2010se}
  \\ \cline{1-3}
     $m_{Bd}$&5.280 [GeV]&\cite{Patrignani:2016xqp}&$h_{A1}(1)$&0.908&\cite{FFC1}\\ 
     $m_{B^-}$&5.279 [GeV]&\cite{Patrignani:2016xqp}&$V_{1}(1)$&1.07&\cite{V10}\\ \cline{4-6}
     $m_{Bs}$&5.367 [GeV]&\cite{Patrignani:2016xqp}& \multicolumn{3}{|c|}{SM particle masses and $G_{\rm F}$}\\ \cline{4-6}
     $M_{Bc}$&6.275 [GeV]&\cite{Patrignani:2016xqp}&$m_\mu$&0.105676 [GeV] &\\ 
     $m_{D}$&1.865 [GeV]&\cite{Patrignani:2016xqp}&$m_\tau$&1.77686 [GeV]&\\
     $m_{D^*}$&2.007 [GeV]&\cite{Patrignani:2016xqp}&$m_c(m_c)$&1.27 [GeV]&\\
     $\tau_{Bd}$&$\hspace{10pt}2.309\times10^{12}$ [GeV$^{-1}$]$\hspace{10pt}$&\cite{Patrignani:2016xqp} &  $m_t$&173.21 [GeV]&\\
     $f_{Bd}\sqrt{B_{Bd}}$ &227.7 [MeV]&\cite{MILC} &      $m_d(2{\rm GeV})$&0.0047 [GeV]&\\
     $\tau_{B^-}$&$\hspace{10pt}2.489\times10^{12}$ [GeV$^{-1}$]$\hspace{10pt}$&\cite{Patrignani:2016xqp} &   $m_s(2{\rm GeV})$&0.096 [GeV]&\cite{Patrignani:2016xqp}\\
     $f_{B^-}$ &186 [MeV]&\cite{HPQCD1} &     $m_b(m_b)$&4.18 [GeV]&\\
      $\tau_{Bs}$& $\hspace{10pt}$$2.294\times10^{12}$ [GeV$^{-1}$]$\hspace{10pt}$&\cite{Patrignani:2016xqp} &      $m_W$&80.385 [GeV]&      \\
     $f_{Bs}\sqrt{B_{Bs}}$ &274.6 [MeV] &\cite{MILC} &      $m_Z$&91.188 [GeV]&\\
     $\tau_{Bc}$&$\hspace{10pt}$$7.703\times10^{11}$ [GeV$^{-1}$]$\hspace{10pt}$&\cite{Patrignani:2016xqp} & $m_h$&125.09 [GeV]&\\
     $f_{Bc}$&0.434 [GeV]&\cite{HPQCD2} & $G_F$&$\hspace{5pt}1.166\times 10^{-5}$ [GeV$^{-2}]\hspace{5pt}$&\\
     \hline
         \end{tabular}
   \end{center}
\end{table}

\end{document}